\documentclass{aa}  

\usepackage{graphicx,url,twoopt}
\usepackage{txfonts}
\usepackage{lipsum}
\usepackage{subcaption}         
\usepackage{lscape}             
\usepackage{placeins}           

\usepackage{longtable}
\usepackage{xcolor}  
\usepackage{float}

\makeatletter
\newcommand{\bibnote}[2]{\@namedef{#1note}{#2}}
\newcommand{\biblink}[2]{\@namedef{#1link}{#2}}
\makeatother

\usepackage{hyperref} 
\usepackage{arydshln}  
\begin{document}

\titlerunning{HST observations of UV lines in R~Leo}
\authorrunning{Saberi et al.}

\title{HST observations of chromospheric UV lines in the AGB star R~Leo}
      
\subtitle{I. C\,\textsc{ii}] multiplet and Mg\,\textsc{ii} h \& k lines}

%
%
%

\author{Maryam Saberi, \inst{\ref{1},\ref{2}}
Donald G. Luttermoser, \inst{\ref{3}}
Graham M. Harper, \inst{\ref{4}}
Theo Khouri, \inst{\ref{5}}
Han Uitenbroek, \inst{\ref{6}}
Wouter Vlemmings \inst{\ref{5}}
      }
\institute{ 
Rosseland Centre for Solar Physics, University of Oslo, P.O. Box 1029 Blindern, NO-0315 Oslo, Norway \label{1}
\and
Institute of Theoretical Astrophysics, University of Oslo, P.O. Box 1029 Blindern, NO-0315 Oslo, Norway \label{2}
\\\email{maryam.saberi@astro.uio.no}
\and
Department of Physics and Astronomy, East Tennessee State University, Johnson City, TN 37614, USA \label{3}
\and 
Center for Astrophysics and Space Astronomy, University of Colorado Boulder, Colorado, CO 80309, USA \label{4}
\and
Department of Physics and Astronomy, Chalmers University of Technology, SE-412 96 Gothenburg, Sweden
\label{5}
\and
National Solar Observatory, Boulder, CO 80303, USA \label{6}
}
\date{}

\abstract {

The role of stellar chromospheres in the chemistry, mass loss, and evolution of cool evolved stars remains poorly understood. We present high-resolution ultraviolet spectra of the nearby Mira-type AGB star R~Leo obtained with STIS on board the \textit{Hubble Space Telescope}. We focus on two strong chromospheric diagnostics, the Mg\,\textsc{ii} h \& k resonance lines and the C\,\textsc{ii}] 2325\,\AA\ multiplet, and model their formation with the NLTE radiative-transfer code RH using phase-dependent hydrodynamic atmospheric structures that include pulsation-driven shocks. The observed C\,\textsc{ii}] multiplet ratios imply an electron density of order $10^9\,\mathrm{cm^{-3}}$. The RH contribution functions show that the C\,\textsc{ii}] emission is strongly localized in a compact shock-heated shell at the first temperature peak of the model atmosphere, near $R\simeq1.8~R_{\rm phot}$, where $T\sim10^4$\,K and the local electron density is a few $10^8\,\mathrm{cm^{-3}}$. In contrast, the Mg\,\textsc{ii} h \& k lines probe a more extended region ($R< 16~R_{\rm phot}$), with the line cores forming at greater radial distances than the wings. The Mg\,\textsc{ii} h line is reproduced more reliably than the k line, which is more strongly affected by circumstellar and interstellar absorption.
As an additional low-opacity kinematic check, the semi-forbidden Al\,\textsc{ii}] $\lambda2669$ line shows a projected stellar-rest-frame blueshift of $\sim6~{\rm km~s^{-1}}$, consistent with shock-related motions expected in Mira atmospheres. Overall, the compact C\,\textsc{ii}] formation region, the extended Mg\,\textsc{ii} contribution functions, and the strong phase dependence of the synthetic Mg\,\textsc{ii} profiles support a picture in which pulsation-driven shocks shape the ultraviolet chromospheric emission lines in Mira variable stars.

}

\maketitle

\section{Introduction}\label{Introduction}

An important long-standing question in stellar astrophysics is understanding processes responsible for heating stellar chromosphere and how these regions would evolve during stellar evolution from the main sequence to red giant branch (RGB) and asymptotic giant branch (AGB) phases. 
The chromosphere in stellar atmospheres is observationally defined as a region above the photosphere where the temperature increases outward. Historically, the chromosphere was not defined by the physical mechanisms responsible for heating, but rather by this observed temperature rise above the photospheric temperature minimum. Later, processes such as acoustic waves, shocks, magnetic activity, pulsations were proposed as mechanisms that transport energy from the convective layers into the upper atmosphere and heat stellar chromospheres \citep[e.g.][]{Linsky1980,Linsky17,Freytag2017,Freytag2023}.

Chromospheres in cool stars are typically revealed through ultraviolet (UV) and optical emission lines formed in the outer atmosphere, including transitions of neutral and singly ionized metals \citep[e.g.][]{Luttermoser1989}. The first evidence for chromospheres in evolved stars was reported by \citep{Herzberg1948, Bidelman1963,Boesgaard1976}, who identified Fe\,\textsc{ii} emission lines in observed spectra of giant stars using McDonald and Mauna Kea telescopes. Later, more UV  spectroscopy observations revealed emission lines such as Mg\,\textsc{ii}, C\,\textsc{ii}, and Fe\,\textsc{ii} are common in the spectra of giants, super giants and asymptotic giant branch (AGB) stars using {\it IAU} and {\it HST} \citep[e.g.][]{Johnson1983, Querci1985, Johnson1986, Johnson1987, Luttermoser1989,Luttermoser1994, Carpenter1997, Wood2000, Rau2018, Dupree2020, Dupree2026}. These UV diagnostics suggest that chromospheres remain present even in these highly evolved stars. The formation of such chromospheric emission lines in cool giants, governed by non-LTE radiative processes, has been discussed by \citep{Judge1990}.

Furthermore, phase-dependent velocity measurements of chromospheric emission lines show that the outer atmospheres of AGB stars are not in hydrostatic and radiative equilibrium, but are instead dominated by dynamic processes such as pulsation-driven shock waves \citep{Willson1976,Willson1982}. Emission lines observed at visual wavelengths, including Fe\,\textsc{i} and the hydrogen Balmer lines, vary in flux over the pulsation cycle \citep{Gillet1988}. Observations with {\it IUE} also showed that UV emission-line variability is correlated with pulsation phase \citep{Bookbinder1989,Luttermoser1996, Wood2000}. 
A recent re-analysis of near-UV spectra of AGB stars by \citet{Guerrero2020} confirmed that the Mg\,\textsc{ii} $\lambda2800$ doublet in regularly pulsating AGB stars is strongly phase dependent.
Consistent with this picture, \citet{Bowen1988} showed that dynamical models of Mira atmospheres predict an extended region of enhanced temperature at a few stellar radii above the photosphere throughout much of the pulsation cycle. This region has been referred to as a {\it calorisphere} by \citet{Willson1986} and as a {\it hydrodynamic chromosphere} by \citet{Dupree1990}, since it resembles a classical chromosphere with $T_{\rm eff} < T < 10^4$ K.

More recently, ALMA observations have resolved the extended atmospheres of nearby AGB stars on sub-stellar-radius scales and revealed that they are highly asymmetric, variable, and dynamically structured. In W Hya, these observations show very hot compact hotspots, warm molecular gas, and simultaneous infall and outflow \citep{Vlemmings2017}. Such high-temperatures gas are thought to arise from hot gas embedded within the cool extended atmosphere. In addition, multi-epoch observations of R~Leo over a period of two weeks showed an outward motion of the millimetre-wavelength surface at a velocity of about 11~km\,s$^{-1}$, likely caused by a shock wave \citep{Vlemmings2019}. 

Despite clear evidence for atmospheric dynamics, the physical origin of UV emission lines in late-type variable stars has remained debated for decades. Proposed mechanisms include pulsation-driven shocks, chromospheric heating, collisional excitation, and radiative or fluorescent pumping \citep[e.g.,][]{Willson1976,Judge1991,JudgeCuntz1993,Judge1993,Ortiz2019,Guerrero2020}. Historically, two broad interpretations have been discussed. In the first, the emission arises in chromospheric layers above the photosphere, where a temperature rise is included in a semi-empirical, quasi-static atmospheric structure; much of the early work based on {\it IUE} ultraviolet spectra followed this approach \citep[e.g.,][]{Johnson1987,Luttermoser1989,Luttermoser1994}. In the second, the emission is associated with outward-propagating shocks driven by stellar pulsation, which heat and compress the gas and naturally produce phase-dependent optical and UV emission features \citep[e.g.,][]{Willson1976,Wood2000,Richter2001}. Although shock-based models have been used to interpret optical emission lines in Mira variables, detailed UV line formation calculations based on time-dependent dynamical AGB atmospheres remain limited.

In this work, we use the phase-dependent atmospheric models of \citet{Bowen1988} to investigate the origin of UV emission lines in the Mira variable R Leo using high-resolution spectra obtained with the Space Telescope Imaging Spectrograph (STIS). Based on these hydrodynamic atmospheric structures, we compute non-LTE synthetic spectra and compare them with the observed STIS data. 
The STIS observations of R Leo and an initial identification of the detected lines were previously presented by \cite{Luttermoser2009}. In the present work, we re-present the observations in order to provide the first detailed analysis of the Mg\,\textsc{ii} and C\,\textsc{ii}] emission lines based on non-LTE modeling with phase-dependent hydrodynamic atmospheres. A forthcoming paper will present a detailed study of Fe\,\textsc{ii} and the remaining detected species.

The paper is organised as follows. In Sect. \ref{Observations}, we present the observations. Section \ref{UV Line Diagnostics} describes the UV diagnostic lines analysed in this study. In Sect. \ref{RH}, we outline the radiative transfer model that we used and the input parameters adopted. The results are presented in Sect. \ref{Results}, and the main conclusions are summarised in Sect. \ref{Summary}.

\begin{table}[]
\caption{Physical properties of the M-type (O-rich) AGB star R Leo.}
\centering
\setlength{\tabcolsep}{2.2pt}
\begin{tabular}{ccccccccc}
\hline
$M_\star$ & $V$ range & $d$ & $\dot{M}$ & $V_{\rm CoM,helio}$ & $R_\star$ & $T_\star$ & $L_\star$ & $P$ \\
($M_{\odot}$) & (mag) & (pc) & ($M_{\odot}\,\mathrm{yr}^{-1}$) & (km s$^{-1}$) & ($R_\odot$) & (K) & ($L_{\odot}$) & (days) \\
\hline
1.5 & $5.2$--$10.6$ & 71 & $1.8\times10^{-7}$ & $7.2\pm1$ & 237 & 2900 & 4100 & 312 \\
\hline
\end{tabular}
\label{Source}
\tablefoot{
The stellar mass is estimated from the $^{17}$O/$^{18}$O = $1.26\pm0.4$ ratio from \cite{Hinkle2016}. 
$R_\star$ is taken as half of the smallest near-infrared uniform-disc diameter measured by \cite{Woodruff2009}; for R Leo, we adopt $D_\mathrm{UD}\approx31$ mas at $\phi\approx0.5$, corresponding to $R_\star\approx15.5$ mas. Using the adopted Gaia distance of $d=71$ pc gives $R_\star\approx237\,R_\odot$. 
The Johnson $V$ magnitude varies over $\sim5.2$--$10.6$ mag according to AAVSO photometry. No $V$-band measurement is available at the exact date of our observation; the AAVSO visual estimate at that epoch is $m_{\rm vis}\approx8.8$, corresponding to phase $\phi\approx0.37$. 
$V_{\rm CoM,helio}$ is the average heliocentric center-of-mass velocity inferred from CO and SiO rotational lines \citep{Hinkle1984}. 
The remaining parameters are taken from \cite{Andriantsaralaza2022}.
}
\end{table}

\section{Observations}\label{Observations}                     

R~Leo is a long-period Mira-type variable star with spectral type M7-9e. The observations analysed in this work were obtained at an optical phase of 0.37, which is close to the phase of maximum UV brightness ($\sim$0.42). The other physical properties adopted for the star are summarised in Table~\ref{Source}.

The observations analysed in this work were obtained with the Space Telescope Imaging Spectrograph (STIS) on board the {\it Hubble Space Telescope} ({\it HST}) toward the AGB star R~Leo (Program ID: 7756).
The spectrum was obtained with the STIS NUV-MAMA detector using the $0.2\times0.2$ arcsec aperture and the E230M grating, which provides a resolving power of $R \approx 30{,}000$, corresponding to a spectral resolution of $\Delta\lambda \approx 0.09$\,\AA\ at the Mg\,\textsc{ii} h \& k lines.
The total exposure time was 184 minutes. The spectra covers the wavelength range 2121.2 - 2941.5 $\AA$. These observations offer the highest UV spectral resolution and the broadest UV range ever obtained for an AGB star currently available. Table \ref{table-data} lists all species identified in the spectra, number of transitions detected for each species, and the ionization energies.\cite{Luttermoser2009} identified and marked all transitions across the entire spectrum, see their figure 1. 

We note that all wavelengths quoted in this paper are vacuum wavelengths. Some earlier studies report air wavelengths; at $\lambda \simeq 2325~\text{\AA}$, this convention difference is about $0.7~\text{\AA}$, or $\sim90~\mathrm{km\,s^{-1}}$, and must therefore be accounted for when comparing line identifications or velocities with other works.


\begin{table}[]
\caption{UV lines detected towards R Leo.}
  \centering
  \setlength{\tabcolsep}{1.8pt}
\begin{tabular}{llll}
\hline
Species & Num. & UV Multiplet& Ioniz. energy(eV) \\
\hline
C\,\textsc{ii}] &  4 & 0.01 & 24.4 \\
Mg\,\textsc{i} & 1 & 9(1)? & 7.6\\
Mg\,\textsc{ii} & 5 & 1(2), 2(1), 3(2) & 15 \\
Al\,\textsc{ii} & 1 & 1 & 18.8\\
Si\,\textsc{i} & 1 & 43? & 8.1\\
Si\,\textsc{ii} & 1 & 0.01 & 16.3\\
V\,\textsc{ii} & 1 & 35(1)? & 14.6\\
Cr\,\textsc{ii} & 10 & 6(1)?, 99(1)?,  & 16.5\\
      &    &  8 (2)?,5 (6) \\
Mn\,\textsc{i}& 3 & 8(1)?, 9(2)? & 7.4\\
Mn\,\textsc{ii} & 5 & 1(3), 5(2) & 15.6 \\
Fe\,\textsc{i}& 3 & 44(2), 45(1) & 7.9 \\
Fe\,\textsc{ii} & 130 & 1(13), 2(10), 3(5), 32(3)  & 16.2\\
&    &        33(1), 35(6), 36(3), 60(8), \\
&    &        61(2),62(9), 63(9), 64(6),  \\
&    &        78(3),132(1), 148(2), 158(4),  \\
&    &        159(2), 161(2), 171(2),179(2),  \\
&    &        195(3),196(1),197(1), 207(1), \\
&    &        218(1),230(1), 235(2),263(1), \\
&    &         277(1),294(2), 301(1), 363(1),  \\
&    &        391(4), 399(6), U(2)  \\
Co\,\textsc{i} & 2 & 6(1)?, 56(1)? & 7.9\\
\hline
\end{tabular}
\label{table-data}
\tablefoot{Third column the UV multiplet numbers for each line and the number of detected line for each multiplet inside parenthesis. Question marks denote lines with uncertain identifications.} 
\end{table}

\section{UV Line Diagnostics}\label{UV Line Diagnostics}

\subsection{C\,\textsc{ii}] multiplet}\label{CII}

\begin{figure*}[]
 \centering
\includegraphics[width=0.85\textwidth]{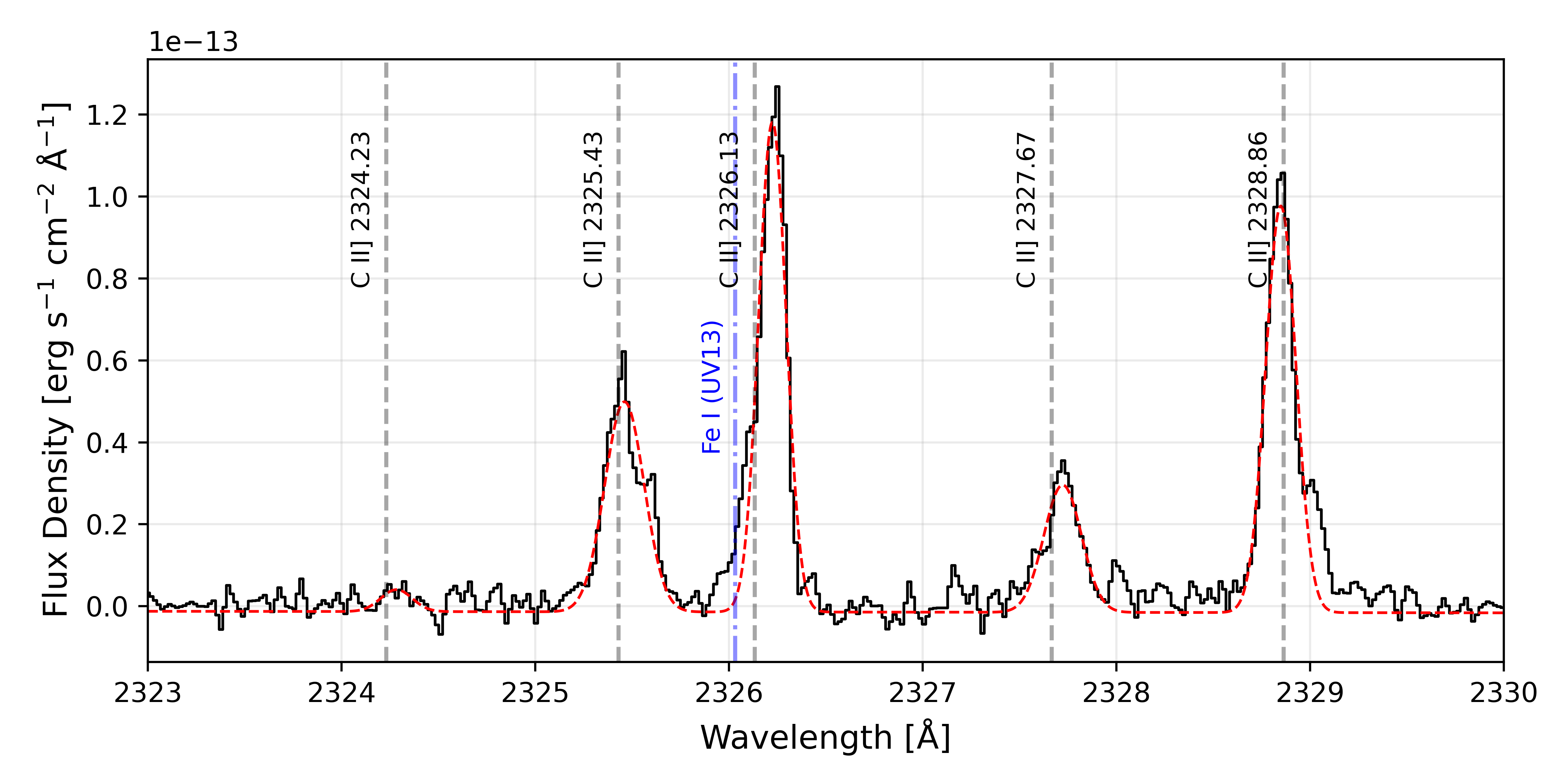}
 \caption[]{C\,\textsc{ii}] 2325\,\AA\ multiplet lines observed toward R~Leo (black solid line), overlaid with the Gaussian fits (red dashed line). The vacuum Ritz wavelengths of the five C\,\textsc{ii}] components are indicated by gray dashed vertical lines. The position of the Fe\,\textsc{ii} UV13 line at 2326.03 \AA, which likely contaminates the C\,\textsc{ii}] 2326.13 \AA\ component, is marked with a blue dash--dot vertical line.}
\label{CII-fig}  
\end{figure*}

In the STIS spectrum, four of the five components of the C\,\textsc{ii}] $\lambda2325$ \AA\ multiplet are clearly detected, while the fifth component is only tentatively identified. Figure~\ref{CII-fig} presents the observed spectrum together with Gaussian fits to the individual components.
Table\ref{CII-Tabel} lists the vacuum Ritz wavelengths and selected atomic parameters for the five C\,\textsc{ii}] $\lambda2325$\,\AA\ multiplet components, taken from the National Institute of Standards and Technology Atomic Spectra Database (NIST ASD (ver. 5.12); \citealt{NIST_ASD}).

The UV~0.01 intercombination multiplet of C\,\textsc{ii}] is the strongest semi-forbidden C$^{+}$ feature in the near-UV and is commonly referred to as the C\,\textsc{ii}] $\lambda2325$ \AA\ multiplet. It consists of five closely spaced components arising from transitions between the upper term $2s,2p^{2},{}^{4}P_{J_u}$ ($J_u=1/2 ,3/2 ,5/2$; $S=3/2$) and the lower term $2s^{2}2p,{}^{2}P^{\circ}{J_\ell}$ ($J_\ell=1/2 ,3/2$; $S=1/2$). Because these transitions involve $\Delta S\neq 0$, they are semi-forbidden (intercombination) lines. The bracket notation C\,\textsc{ii}] indicates a spin-forbidden electric-dipole transition, which leads to small transition probabilities and correspondingly low radiative decay rates (see low $A_{ul}$ values in Table \ref{CII-Tabel}).


The relatively small radiative transition probabilities of the intercombination C\,\textsc{ii}] UV~0.01 multiplet make the level populations sensitive to the competition between radiative decay and collisional excitation and de-excitation. As a result, the relative intensities of the C\,\textsc{ii}] lines near $2325\,\text{\AA}$ provide a well-established diagnostic of the electron density in cool-star chromospheres. Unlike strong resonance lines such as Mg\,\textsc{ii} h \& k, which are often optically thick and require partial-redistribution radiative transfer, the C\,\textsc{ii}] intercombination lines are generally much less optically thick and are commonly treated as collisionally excited lines followed by radiative decay. Their flux ratios are therefore valuable probes of the electron density in the upper chromospheres of late-type stars \citep[e.g.,][]{Stencel1981,Luttermoser1989,Judge1994,Harper2006}.
In a pioneer study, \citet{Stencel1981} investigated the density sensitivity of the boron-like C\,\textsc{ii}] intercombination multiplet for chromospheric conditions in the Sun and in cool evolved stars observed with the ({\it IUE}). They showed that flux ratios within the multiplet are most sensitive over electron densities of approximately $10^{7} \lesssim n_e \lesssim 10^{9} ~\mathrm{cm^{-3}}$, matching the typical values observed for evolved late-type stars, while the solar chromosphere lies close to the high-density limit ($n_e \sim 10^{11} ~\mathrm{cm^{-3}}$). This makes the C\,\textsc{ii}] multiplet one of the most useful electron-density diagnostics for stellar chromospheres in cool stars. An additional advantage is that the ratios are only weakly dependent on temperature, since the collisional de-excitation rates scale approximately as $T_e^{-1/2}$, reducing the sensitivity of the inferred $n_e$ to uncertainties in $T_e$ \citep{Harper2006}. 

\begin{table*}[t]
  \setlength{\tabcolsep}{5pt}
    \caption{C\,\textsc{ii}] multiplet line parameters along with the Gaussian fitting parameters from Fig. \ref{CII-fig}.}
  \begin{tabular}{ccllcccccccc}
\hline
Wavelength & Multiplet & $A_{ul}$ & $E_l$ & $E_u$ & $g_l$ & $g_u$ & $v_{\rm obs}$ & $v_*$ & Peak & FWHM & $F_\nu$ \\
(${\AA}$) & UV & (s$^{-1}$) & (cm$^{-1}$) & (cm$^{-1}$) & & & (km s$^{-1}$) & (km s$^{-1}$) & [erg\,s$^{-1}$\,cm$^{-2}$\,\AA$^{-1}$] & (\AA/km\,s$^{-1}$) & (ergs s$^{-1} cm^{-2}$) \\
\hline
2324.236 & 0.01 & 1.4  & 0.00  & 43024.9 & 2 & 4 & 6.94  & -0.26 & $5.35\times10^{-15}$ & 0.200 / 25.84 & $1.14\times10^{-15}$\\
2325.430 & 0.01 & 59.9 & 0.00  & 43002.8 & 2 & 2 & 3.97  & -3.23 & $5.13\times10^{-14}$ & 0.259 / 33.43 & $1.41\times10^{-14}$\\
2326.133 & 0.01 & 44.3 & 63.39 & 43053.2 & 4 & 6 & 11.78 &  4.58 & $1.19\times10^{-13}$ & 0.173 / 22.34 & $2.20\times10^{-14}$ \\
2327.665 & 0.01 & 8.47 & 63.39 & 43024.9 & 4 & 4 & 7.46  &  0.26 & $3.11\times10^{-14}$ & 0.244 / 31.48 & $8.08\times10^{-15}$\\
2328.863 & 0.01 & 67.8 & 63.39 & 43002.8 & 4 & 2 & -1.82 & -9.02 & $9.92\times10^{-14}$ & 0.193 / 24.88 & $2.04\times10^{-14}$ \\
\hline
 \end{tabular}
\tablefoot{Spectroscopic data are taken from the NIST-ASD. The listed wavelengths are Ritz vacuum wavelengths. Integrated fluxes, full widths at half maximum (FWHM), and observed line-center velocity offsets ($v_{\rm obs}$) are derived from Gaussian fits to the observed line profiles. The quantity $v_{\rm obs}$ is measured relative to the corresponding Ritz vacuum wavelength. The stellar-rest-frame velocity is computed as $v_* = v_{\rm obs} - V_{\rm CoM}$, adopting $V_{\rm CoM}=7.2$ km s$^{-1}$.}
 \label{CII-Tabel}
\end{table*}

\subsection{Mg\,\textsc{ii} h $\&$ k lines}\label{MgII}

In the STIS spectra five Mg\,\textsc{ii} transitions are detected. The atomic data and Gaussian fitting parameters for all transitions are listed in Table~\ref{MgII-table}. The atomic data are taken from the NIST-ASD. All wavelengths listed in all tables are calculated Ritz wavelengths in vacuum.

The Mg\,\textsc{ii}~h \& k resonance lines at 2803 and 2796\,\AA\ are among the strongest UV diagnostics of chromospheres in cool stars \citep[e.g.][]{Linsky1980}. These lines arise from transitions between the ground state and the first excited levels of singly ionized magnesium ($3s\,^2S_{1/2} \rightarrow 3p\,^2P_{1/2,3/2}$). Because magnesium is easily ionized in the upper photosphere and chromosphere of cool stars, Mg\,\textsc{ii} is abundant in these layers, making the Mg\,\textsc{ii}~h \& k lines extremely optically thick. Their formation is dominated by resonance scattering and collisional excitation and therefore requires non-LTE radiative transfer for accurate modeling \citep{Mihalas1978,Wood2000ApJ, Hubeny2014}. Solar chromospheric models indicate that the wings of the Mg\,\textsc{ii}~h \& k lines form in the upper photosphere and lower chromosphere, while the emission peaks and central core originate higher in the chromosphere \citep{Vernazza1981}. A similar qualitative formation structure is expected in the chromospheres of other cool stars. As a result, Mg\,\textsc{ii}~h \& k line profiles probe a broad range of atmospheric heights and are sensitive to the chromospheric temperature structure, velocity fields, and dynamic phenomena such as shocks and mass outflows. In pulsating evolved stars, these lines provide important diagnostics of shock propagation and chromospheric heating in the extended atmospheres of Mira and semiregular variables \citep{Luttermoser2009}.

In Mira stars, the Mg\,\textsc{ii} k line is often strongly affected by overlying circumstellar absorption from nearby Fe\,\textsc{i} and Mn\,\textsc{i} lines. In particular, Fe\,\textsc{i} at $\lambda_{\rm air}=2795.005$~\AA\ ($\lambda_{\rm vac}\approx2795.829$~\AA) and Mn\,\textsc{i} at $\lambda_{\rm air}=2794.816$~\AA\ ($\lambda_{\rm vac}\approx2795.640$~\AA) can obscure a significant fraction of the emitted Mg\,\textsc{ii} k flux \citep{Luttermoser1989,Luttermoser2000,Wood2000}. In contrast, the Mg\,\textsc{ii} h line is generally less affected by these circumstellar absorption features and therefore provides a clearer probe of the chromospheric emission.

The velocity offsets listed in Table \ref{MgII-table} are measured relative to the Ritz vacuum wavelengths and are reported both in the observed frame ($v_{\rm obs}$) and in the stellar rest frame ($v_\star$). Both Mg,\textsc{ii} resonance lines are blueshifted in the stellar rest frame, with $v_\star=-28.7,{\rm km,s^{-1}}$ for Mg,\textsc{ii} k and $v_\star=-38.1,{\rm km,s^{-1}}$ for Mg,\textsc{ii} h. However, following \citet{Wood2000,Wood2000ApJ}, we do not interpret these shifts as direct measurements of the bulk gas velocity, since the Mg,\textsc{ii} h \& k lines are optically thick and their centroids are strongly influenced by scattering and opacity effects. The observed blueshifts instead indicate that the escaping Mg,\textsc{ii} photons preferentially emerge from the blue side of the line profile.

\begin{table*}[t]
  \setlength{\tabcolsep}{3.5pt}
    \caption{Detected Mg\,\textsc{ii} lines towards R Leo.}
  \begin{tabular}{@{} ccllccccccccccc@{}}
\hline
Wavelength & Multiplet & $A_{ul}$ & $E_l$ & $E_u$ & $g_l$ & $g_u$
& $v_{\rm obs}$ & $v_*$ & Peak & FWHM & $F_\nu$ \\
(${\AA}$) & UV & ($\rm 10^{8}~ s^{-1}$) & (cm$^{-1}$) & (cm$^{-1}$)
& & & (km s$^{-1}$) & (km s$^{-1}$)
& [erg\,s$^{-1}$\,cm$^{-2}$\,\AA$^{-1}$]
& (\AA/km\,s$^{-1}$)
& (ergs s$^{-1} cm^{-2}$) \\
\hline
2796.352 & 1 (k) & 2.60 & 0.00 & 35760.9 & 2 & 4 & -21.5 & -28.7 & $7.29\times10^{-12}$ & 0.226 / 24.2 & $1.75\times10^{-12}$\\

2803.531 & 1 (h) & 2.57 & 0.00 & 35669.3 & 2 & 2 & -30.9 & -38.1 &
$8.71\times10^{-12}$ & 0.416 / 44.5 & $3.86\times10^{-12}$\\

\hline
2791.600 & 3 & 4.01 & 35669.3 & 71491.1 & 2 & 4 &
-3.5 & -10.7 &
$9.49\times10^{-14}$ & 0.227 / 24.4 & $2.29\times10^{-14}$\\

2798.823 & 3 & 4.79 & 35760.9 & 71490.2 & 4 & 6 &
-0.7 & -7.9 &
$3.57\times10^{-14}$ & 0.223 / 23.9 & $8.47\times10^{-15}$\\

2929.490 & 2 & 1.15 & 35669.3 & 69804.9 & 2 & 2 &
0.1 & -7.1 &
$1.79\times10^{-14}$ & 0.289 / 29.6 & $5.53\times10^{-15}$\\
\hline
 \end{tabular}
\tablefoot{Spectroscopic data are taken from the NIST-ASD. The listed wavelengths are Ritz vacuum wavelengths. Integrated fluxes, full widths at half maximum (FWHM), and observed line-center velocity offsets ($v_{\rm obs}$) are derived from Gaussian fits to the observed profiles. The stellar-rest-frame velocity is computed as $v_* = v_{\rm obs} - V_{\rm CoM}$, adopting $V_{\rm CoM,helio}=+7.2~{\rm km~s^{-1}}$. We note that the FWHM listed for Mg\,\textsc{ii} k was measured from a Gaussian fit to the stronger red component only, rather than to the full double-peaked profile. Therefore, the tabulated k-line FWHM should not be interpreted as the intrinsic width of the full Mg\,\textsc{ii} k emission feature.}
\label{MgII-table}
\end{table*}

\section{RH Radiative Transfer Calculations}\label{RH}

To simulate the Mg\,\textsc{ii} h \& k lines and the C\,\textsc{ii}] multiplet, we used the non-LTE radiative-transfer code RH \citep{Uitenbroek2001}. RH is based on the MALI (Multi-level Approximate Lambda Iteration) formalism of \citet{Rybicki1991,Rybicki1992} and solves the coupled radiative-transfer and statistical-equilibrium equations for multi-level atoms under general NLTE conditions.

The atmospheric stratifications (temperature, mass density, velocity, and column mass) were taken from the dynamical models of \citet{Bowen1988} at the selected pulsation phases and used as input to RH, as described in Sect.~\ref{ModelAtmosphere}. The electron density was calculated internally by RH using the \texttt{SOLVE=ONCE} option. In this mode, RH solves the charge conservation equation once at the beginning of the calculation and then keeps the resulting electron density fixed during the subsequent NLTE iterations. The charge-conservation calculation includes electron contributions from metals treated in LTE, as well as from the atoms included in the RH atomic setup. For the active atoms, the charge contributions associated with the NLTE atomic populations are accounted for in the charge balance. In addition, the hydrogen populations supplied in the input atmosphere are read into the background atomic hydrogen population structure, so the electron contribution associated with hydrogen ionization is included. Thus, the electron density used in the radiative-transfer calculation includes the hydrogen contribution and is not a metal-only LTE estimate. 

The calculations were performed with the one-dimensional version of the code. In the treatment of scattering and the line source function, RH can employ either complete redistribution (CRD) or partial redistribution (PRD). CRD assumes that absorbed and emitted photon frequencies are uncorrelated, whereas PRD accounts for their partial correlation during scattering. For the
Mg\,\textsc{ii} lines we used PRD, which is essential for modelling strong resonance lines such as Mg\,\textsc{ii} h \& k and strongly affects their cores and inner wings. Although the input atmospheric structures are spherically extended, PRD calculations with the spherical implementation of RH did not converge for the Mg\,\textsc{ii} h \& k lines; therefore, we employed the 1D
version in order to retain the PRD treatment.

The RH atomic setup includes the species H, He, C, Mg, Fe, Na, O, Si, Al, Ca, and S. Solar abundances were adopted from Grevesse \& Anders (1991), except for carbon, oxygen, and iron, for which the updated abundances $A(\mathrm{C}) = 8.39$, $A(\mathrm{O}) = 8.66$, and $A(\mathrm{Fe}) = 7.44$ were used, following \citep{Asplund2005, Asplund2004, Asplund2000}, respectively.
Among these, H, Mg, and C are treated as active atoms and are solved in full NLTE, meaning that their level populations are obtained by solving the coupled radiative-transfer and statistical-equilibrium equations iteratively. The remaining species are
included as passive background atoms, contributing to the background opacity and charge balance but not to the full NLTE rate-equation iterations. This setup allows the models to converge within a reasonable computational time while retaining a full NLTE treatment for the species analysed in this work.

\subsection{Model Atoms}\label{ModelAtoms} 


The hydrogen atom is represented by a five-bound-level-plus-continuum model consisting of five H\,\textsc{i} bound levels ($n=1$--5) and the H\,\textsc{ii} continuum. The model includes ten bound--bound radiative transitions spanning the Lyman, Balmer, Paschen, and Brackett series, together with five bound--free continua treated using hydrogenic photoionization cross sections. Partial frequency redistribution (PRD) is adopted for Ly$\alpha$ and Ly$\beta$. Although simplified, this hydrogen atom model includes the main bound--bound and bound--free processes needed for the non-LTE calculation,, while allowing the models to converge in a reasonable computational time.


Magnesium is also represented by a simple four-level atom consisting of the Mg\,\textsc{ii} ground state, the two fine-structure upper levels of the resonance doublet, and the Mg\,\textsc{iii} ground state. This model is intended specifically to describe the formation of the Mg\,\textsc{ii} h and k lines, which are treated with partial frequency redistribution (PRD). Although simplified, it retains the essential atomic structure needed for the non-LTE calculation of the strong Mg\,\textsc{ii} resonance emission analysed in this work.

Carbon is represented by a multi-level atom that includes 23 levels spanning C\,\textsc{i}, C\,\textsc{ii}, and the C\,\textsc{iii} ground state. This model is designed to describe both the ionization balance of carbon and the main transitions relevant for the UV spectrum. In particular, it includes the low-lying fine-structure levels of C\,\textsc{ii} that give rise to the C\,\textsc{ii}] 2325\,\AA\ multiplet analysed in this work. The model therefore provides the basic non-LTE treatment needed to compute the C\,\textsc{ii}] emission while keeping the atomic setup simple enough for efficient calculations.

\subsection{Model Atmosphere}\label{ModelAtmosphere} 

%


We use atmospheric structures from the dynamical models of Bowen (1988), kindly provided by G.~H.~Bowen. 
The Bowen (1988) models are one-dimensional, spherically symmetric, time-dependent hydrodynamic simulations in which the stellar pulsation is represented by a sinusoidally moving inner boundary. The temperature structure is computed assuming LTE grey radiative transfer, and dust formation is treated with a simplified time-dependent prescription that enables radiation pressure on dust to drive the mass loss. 
The high temperatures reached in the inner atmosphere are produced self-consistently in the post-shock regions and do not result from any chromospheric heating term, which is not included in the models. 
We adopt these models because they provide phase-dependent atmospheric stratifications with shock-heated layers appropriate for R~Leo-like conditions and sufficiently high temperatures for testing whether the observed UV emission can be produced by the dynamical atmosphere itself, without imposing an additional semi-empirical chromosphere.

The model used in this paper is calculated for a Mira-type star at eight phases covering a full pulsation cycle. The model parameters are given in Table~\ref{Bowen-T}. The effective temperature and pulsation period are slightly higher than those of R~Leo (Table~\ref{Source}); however, the differences are small enough that the model provides a reasonable approximation for our purposes. 

We assumed $\log g = -0.32$ (cgs) using the initial mass and radius from Table~\ref{Bowen-T}.
The equilibrium (static) stellar radius used to construct the initial model is 239.40\,$R_\odot$. The piston velocity amplitude is 3\,km\,s$^{-1}$, which drives the periodic shocks that propagate through the atmosphere and produce strong phase-dependent variations in the thermal and dynamical structure. The optical phases and the corresponding photospheric radii are given in Table~\ref{Bowen-photosphere}. The photospheric radius varies by about 20\% over the pulsation cycle, reflecting the expansion and contraction of the stellar atmosphere, and this variation is taken into account when scaling the radial stratifications for each phase.

\begin{figure}[]
 \centering
\includegraphics[width=0.5\textwidth]{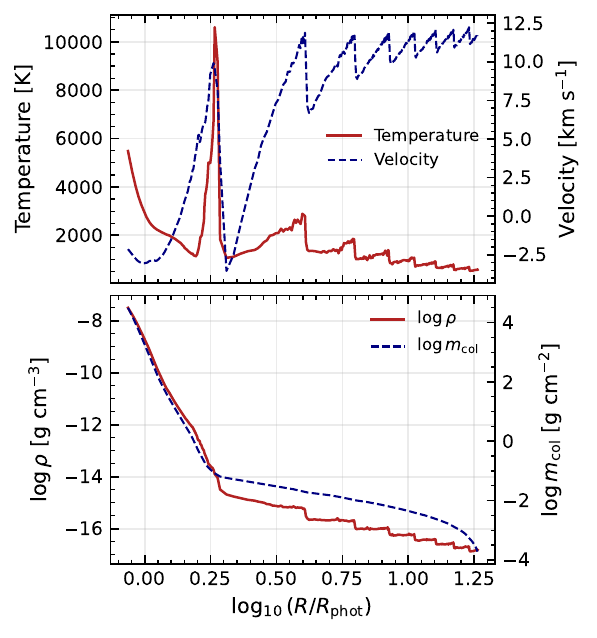}
 \caption[]{Temperature, velocity, gas density and column mass density as a function of radius from the Bowen dynamical model at the optical phases 0.377. Microturbulence velocity is assumed 3 km s $^{-1}$ in all models. This atmospheric structure is used as input for the RH radiative-transfer modelling of the Mg\,\textsc{ii}~h \& k lines. The photospheric radius is estimated to be $R_{\rm phot}\sim 1.843\times10^{13}$ cm at this phase.}
\label{Bowen-P4}  
\end{figure}

From the Bowen models we obtain stratifications of temperature, gas density and column mass. The atmospheric model for our main model P4 corresponding to $\phi=0.377$ is shown in Fig. \ref{Bowen-P4}.  
This model is used as input for the atmospheric model in the RH radiative-transfer calculations to reproduce the observed Mg\,\textsc{ii} h \& k and C\,\textsc{ii}] lines (Sect.~\ref{Results}). The atmospheric stratifications for all eight phases from Bowen calculations are shown in Fig.~\ref{Fig-Bowen-8phase}.

\begin{table}[]
\caption{Assumed model parameters of Bowen model.}
  \centering
  \setlength{\tabcolsep}{3.5pt}
\begin{tabular}{ccccc}
\hline
$M_\star$  & Piston amplitude & $R_{\star}$ &  $T_{\star}$  & P  \\
\tiny($M_{\odot}$) & \tiny(km s$^{-1}$)& ($R_{\odot}$) & \tiny(K) & \tiny(days) \\
\hline
1.0 & 3.0 & 239.40 & 3000  & 320  \\
\hline
\end{tabular}
\label{Bowen-T}
\end{table}

\section{Results}\label{Results}

\subsection{Electron density estimation from C\,\textsc{ii}] multiplet}\label{CII-Ne-results}

Definitions of the five diagnostic ratios of the C\,\textsc{ii}] multiplet commonly used in studies of the atmospheric structure of cool evolved stars, together with the observed values for R~Leo, are listed in Table \ref{CII-ratio-table}. In principle, ratios comparing lines arising from different upper fine-structure levels (usually denoted $R_1$--$R_3$) are sensitive to the electron density, $n_e$, because collisional coupling and de-excitation modify the relative populations of the $^{4}P_{J_u}$ levels as the density increases. By contrast, ratios between lines sharing the same upper level ($R_4$--$R_5$) approach fixed branching ratios in the optically thin limit and are therefore mainly sensitive to optical-depth effects \citep[e.g.][]{Judge1994,Harper2006}. 

In practice, however, blending can affect individual components and bias some of the ratios. It is therefore important to identify the least contaminated line combinations or, where possible, correct for flux losses when deriving $n_e$.
As shown in Fig. \ref{CII-fig}, the C\,\textsc{ii}] component at 2326.133\,\AA\ appears to be affected by circumstellar absorption from a nearby Fe\,\textsc{i} transition, Fe\,\textsc{i} (UV13) at 2326.03\,\AA, marked by the blue dashed line in Fig. \ref{CII-fig}. This absorption produces a fluorescent Fe\,\textsc{i} (UV45) emission line at $\lambda_{\rm vac}=$2807.811\,\AA, since both Fe\,\textsc{i} transitions share the same upper level ($\mathrm{z}\,^5H^o$, $J=5$) \citep{Luttermoser2000,Luttermoser2009}.
In order to use the C\,\textsc{ii}] 2326.133\,\AA\ line for the electron-density estimate, we estimated its flux lost using the measured flux of the fluorescent Fe\,\textsc{i} (UV45) line that is marked in Fig. 1 in \cite[][]{Luttermoser2009} paper.

From a Gaussian fit to the Fe\,\textsc{i} UV45 line, we obtained an integrated flux of $F_{\mathrm{Fe\,I}} = 2.89\times10^{-14}\ \mathrm{erg\ s^{-1}\ cm^{-2}}$. To estimate the equivalent C\,\textsc{ii}] flux lost from the 2326.133\,\AA\ line through Fe\,\textsc{i} pumping, we scale this flux by the photon-energy ratio,
\[
F_{\mathrm{lost}} = F_{\mathrm{Fe\,I}} \times \frac{2807.811}{2326.133},
\]
which gives $F_{\mathrm{lost}} \approx 3.49\times10^{-14}\ \mathrm{erg\ s^{-1}\ cm^{-2}}$. Using the observed C\,\textsc{ii}] 2326.133\,\AA\ integrated flux from Table \ref{CII-Tabel},
\[
F_{\mathrm{obs}} = 2.2\times10^{-14}\ \mathrm{erg\ s^{-1}\ cm^{-2}},
\]
the corrected intrinsic C\,\textsc{ii}] flux becomes
\[
F_{\mathrm{corr-CII(2326)}} = F_{\mathrm{obs}} + F_{\mathrm{lost}} \approx 5.69\times10^{-14}\ \mathrm{erg\ s^{-1}\ cm^{-2}}.
\]

We corrected R1 and R2 ratios using the $F_{\rm corr-CII(2326)}$ and those are listed as $R_{\rm corr}$ in Table \ref{CII_ratio-table}.

\begin{figure}[]
 \centering
\includegraphics[width=0.5\textwidth]{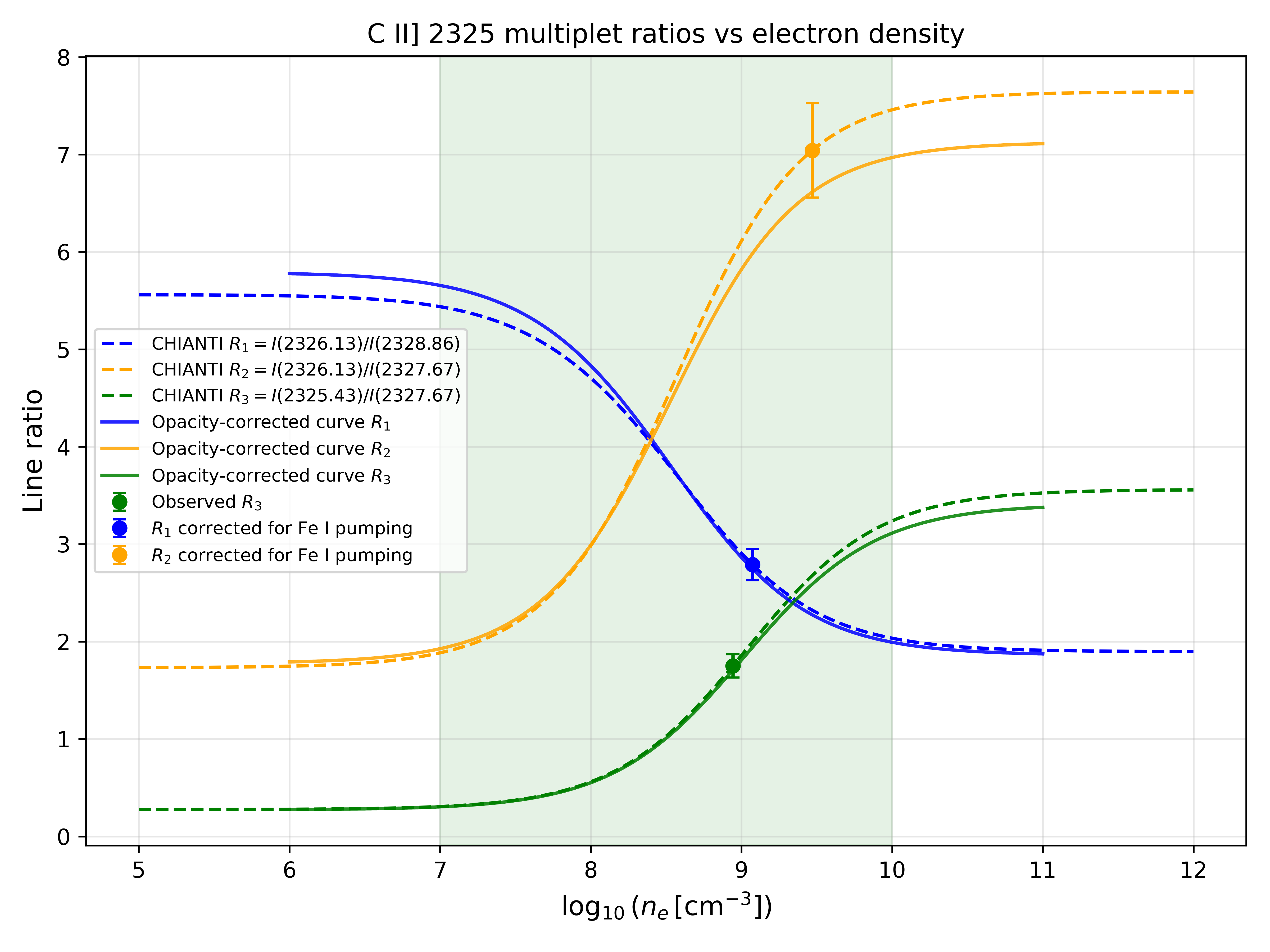}
 \caption[]{Electron-density diagnostic curves for the C\,\textsc{ii}] 2325\,\AA\ intercombination multiplet. The dashed lines show the CHIANTI ratios for $T_e = 7000$\,K, and the solid lines show the opacity-corrected ratios. The observed ratios are shown as markers, with $R_1$ and $R_2$ corrected for the Fe\,\textsc{i} pumping/fluorescence effect.}
\label{Ne-CII}  
\end{figure}

To derive the electron density from $R_1$, $R_2$, and $R_3$, we used the CHIANTI atomic database and software package to compute the C\,\textsc{ii}] 2325\,\AA\ multiplet line ratios as a function of electron density. CHIANTI solves the collisional-radiative statistical-equilibrium equations for the relevant C$^{+}$ energy levels under the optically thin assumption, using atomic data such as energy levels, radiative transition probabilities, and electron-impact excitation rates to predict the line emissivities for a given electron temperature, $T_e$, and electron density, $n_e$. From these emissivities we calculated the diagnostic ratio curves shown in Fig. \ref{Ne-CII}. We tested $T_e=6000$, 7000, 8000, and 10\,000~K and found that the resulting CHIANTI curves change very little over this temperature range. We therefore show only the CHIANTI curves for $T_e=7000$~K here. In addition, we also calculated our own corrected ratio curves to estimate how opacity affects the ratios; these corrected curves are not from CHIANTI.

Figure \ref{Ne-CII} shows the electron-density diagnostic curves.
The dashed curves show the CHIANTI ratios for $T_e=7000$\,K under the optically thin assumption. The solid curves are our own calculations including the effect of opacity on the ratios. The green shaded region marks the density range where the ratios are most sensitive to the electron density $n_e$. The observed ratios, derived from Gaussian-integrated line fluxes, are shown as markers. Ratios involving the 2326.133\,\AA\ line ($R_1$ and $R_2$) are corrected for the flux lost through the Fe\,\textsc{i} pumping/fluorescence process. Among the three ratios, $R_3$ is the cleanest density tracer, since it does not involve the corrected 2326.133\,\AA\ line. It gives $n_e \sim 8.9\times10^8\,\mathrm{cm^{-3}}$, while $R_1$ and $R_2$ are consistent with this value within their uncertainties, although $R_2$ has the largest error bar. Finally, the measured values of $R_4$ and $R_5$ are in good agreement with the expected constant values of 7.28 and 1.04, respectively \citep{Judge1998}.

\begin{table*}
\centering
\caption{C\,\textsc{ii}] $2s^{2}2p\;{}^{2}P^{\circ} - 2s\,2p^{2}\;{}^{4}P$ $n_e$-sensitive and opacity diagnostics.}
\begin{tabular}{lccccc}
\hline\hline
Ratio & Definition (vacuum wavelengths) & Transitions ($J_u \rightarrow J_\ell$) 
      & Diagnostic & $R_{\rm obs}$ & $R_{corr}$ \\
\hline
R1 & $2326.133 / 2328.863$ &
     $(5/2 \rightarrow 3/2)/(1/2 \rightarrow 3/2)$ &
    $n_e$ & $1.08 \pm 0.06$  & $2.79 \pm 0.16$\\

R2 & $2326.133 / 2327.665$ &
     $(5/2 \rightarrow 3/2)/(3/2 \rightarrow 3/2)$ &
     $n_e$ & $2.73 \pm 0.19$  & $7.04 \pm 0.48$\\

R3 & $2325.430 / 2327.665$ &
     $(1/2 \rightarrow 1/2)/(3/2 \rightarrow 3/2)$ &
     $n_e$ & $1.75 \pm 0.12$  & --\\
\hline
R4 & $2327.665 / 2324.236$ &
     $(3/2 \rightarrow 3/2)/(3/2 \rightarrow 1/2)$ &
     $N_{\mathrm{C\,II]}}$ &  $7.09$ & -- \\

R5 & $2328.863 / 2325.430$ &
     $(1/2 \rightarrow 3/2)/(1/2 \rightarrow 1/2)$ &
     $N_{\mathrm{C\,II]}}$ &  $1.44$ & -- \\
\hline
\end{tabular}
\label{CII-ratio-table}
\end{table*}

\subsection{NLTE calculations of Mg\,\textsc{ii} h \& k lines}\label{MgII-results}


We performed calculations with the RH radiative transfer code for all eight Bowen snapshots shown in Fig. \ref{Bowen-photosphere}. The results for all other phases are shown in fig. \ref{Fig-MgII-8phase}.
Here we focus on the model calculated for P4 which is presented in Fig. \ref{Bowen-P4}; and serve as our main model corresponding to the optical phase of the observations. This model also gave us the best matching results with the observations as expected. N-LTE calculations are solved for Mg, C and H simultaneously.

\begin{figure*}[]
 \centering
\includegraphics[width=0.90\textwidth]{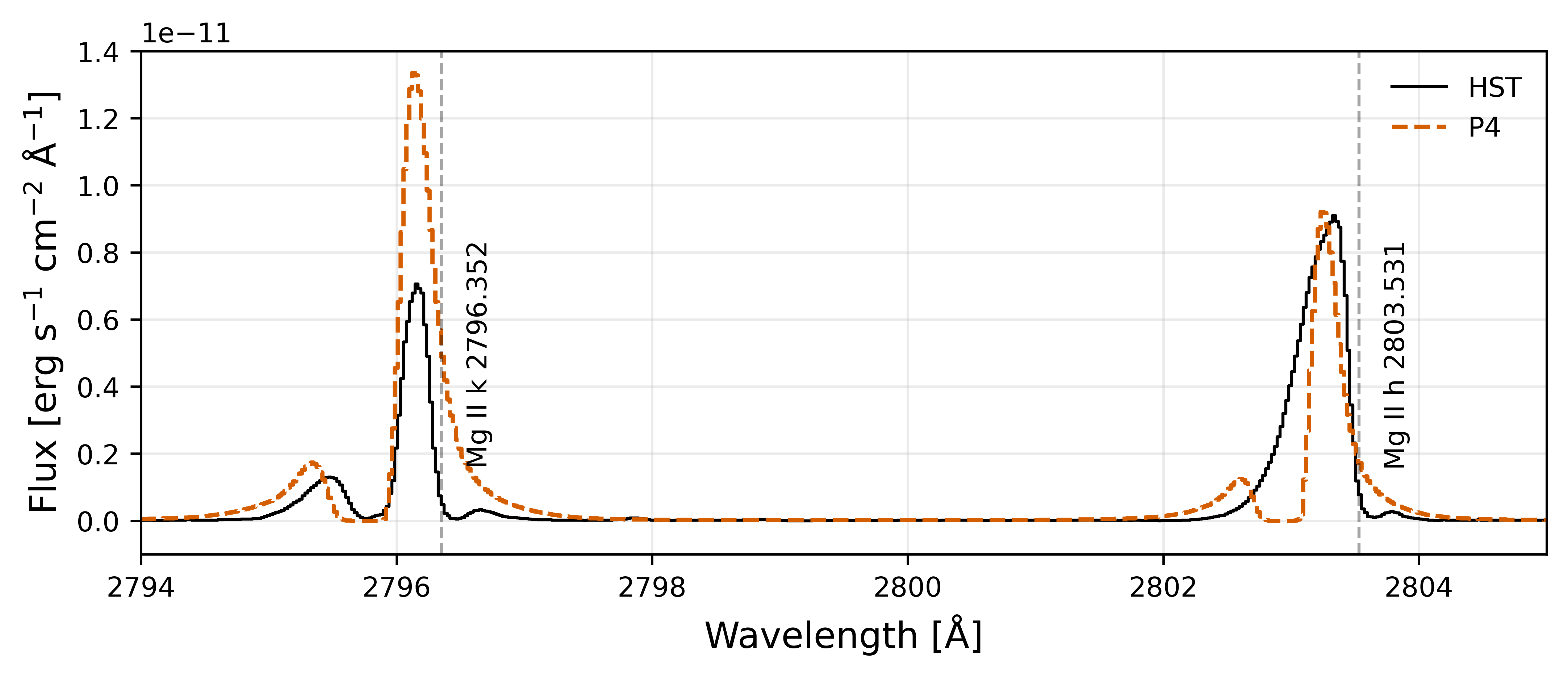}
 \caption[]{STIS spectra of the Mg\,\textsc{ii} $h$ \& $k$ lines (black) compared with the RH model results for P4 ($\phi=0.377$; orange). }
\label{MgII-P4}  
\end{figure*}

The Mg\,\textsc{ii} h \& k lines overlaid with the model spectra is shown in Fig. \ref{MgII-P4}. As can be seen the Mg\,\textsc{ii} k line is over-predicted and the Mg\,\textsc{ii} h line is narrower than the observed profile.
As it was mentioned earlier in Sect. \ref{MgII} the Mg\,\textsc{ii} k line in Mira stars is often strongly affected by circumstellar absorption from Fe\,\textsc{i} (UV3) at $\lambda_{\rm vac}\approx2795.82$\,\AA\ and Mn\,\textsc{i} at ($\lambda_{\rm vac}\approx2795.640$ \AA), which can significantly reduce the observed flux in the $k$ line. This explains the overestimated flux of Mg\,\textsc{ii} k in the RH model compared the observed line. In contrast, the Mg\,\textsc{ii} h line is generally less affected by circumstellar absorption and therefore provides a more reliable diagnostic of the intrinsic chromospheric emission.
The mismatch in the Mg\,\textsc{ii} h-line width may reflect missing macroturbulent broadening or unresolved velocity structure in our one-dimensional model, similar to the issue discussed by \citet{Wood2000ApJ}, who required an additional macroturbulent component to reproduce the observed Mg\,\textsc{ii} profiles of Mira variables.
Additionally, the Mg\,\textsc{ii} h line profile look to be slightly asymmetric, with a red side that is steeper than the blue side. This asymmetry has been seen in the Mg\,\textsc{ii} h spectra of other AGB stars as well \citep{Wood2000}.

To reproduce the observed absolute Mg\,\textsc{ii} fluxes, the emergent RH spectra were scaled by a geometrical dilution factor $(R_{\rm eff}/d)^2$, where $R_{\rm eff}$ is the effective radius of the emitting region. Adopting a distance of 71\,pc and the Bowen-model photospheric radius at P4, $R_{\rm phot}=1.843\times10^{13}$\,cm, corresponding to $R_{\rm phot}\simeq265\,R_\odot$, we obtain a best-fit value of $R_{\rm eff}=18\,R_\odot$. This corresponds to an angular diameter of $2.36$\,mas, compared with a photospheric angular diameter of $34.70$\,mas. We emphasize that ($R_{\rm eff}$) should not be interpreted as the physical radius of the Mg\,\textsc{ii}-forming region. Rather, it represents the effective emitting area required to reproduce the observed absolute flux and is therefore analogous to a filling factor. The contribution-function analysis in Sect. \ref{MgII-formation} indicates that the Mg\,\textsc{ii} emission originates over a much more extended region of the atmosphere.

\subsection{Mg\,\textsc{ii} h \& k formation regions }\label{MgII-formation}

\begin{figure*}
  \centering
  \includegraphics[width=0.9\textwidth]{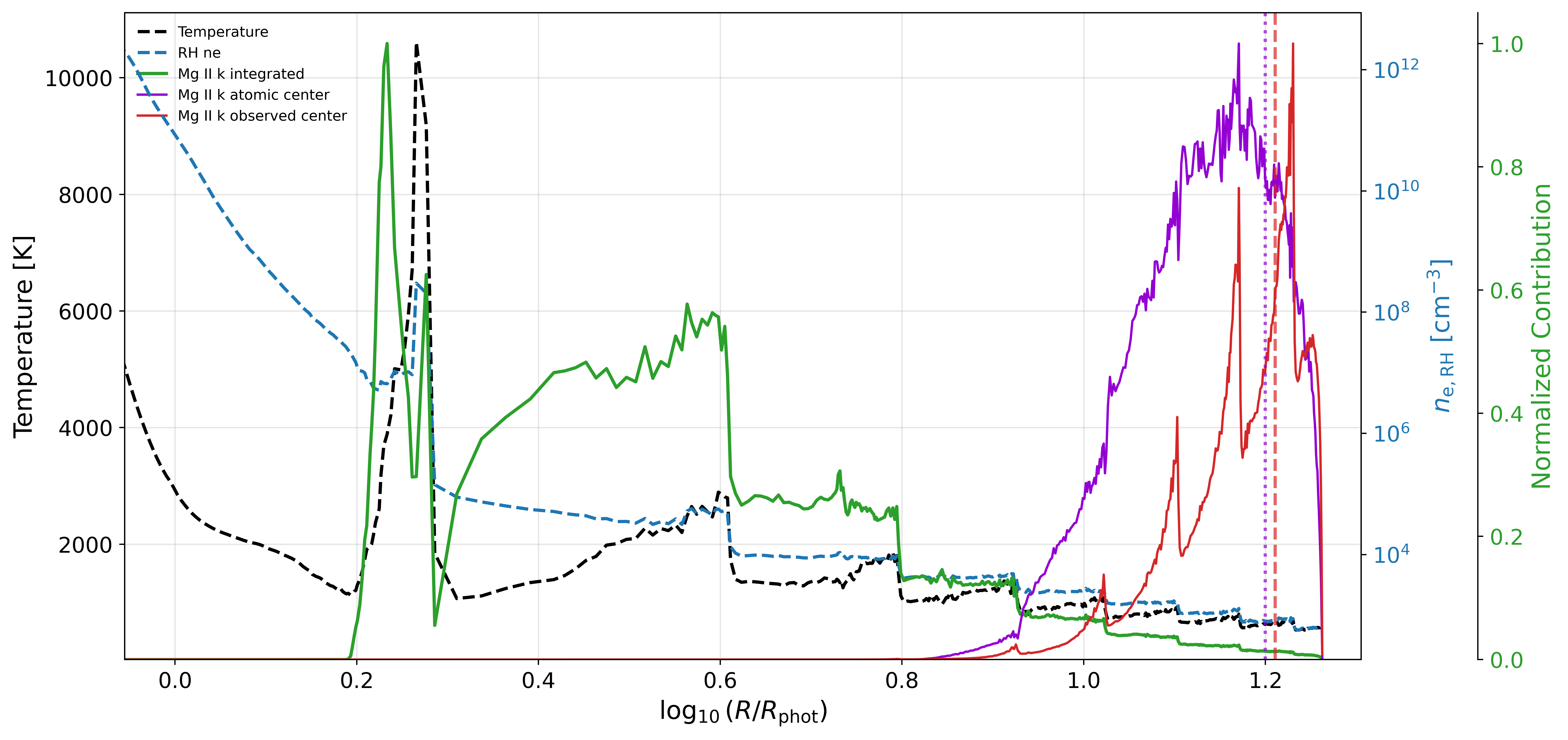}
  \vspace{0.5cm}
  \includegraphics[width=0.9\textwidth]{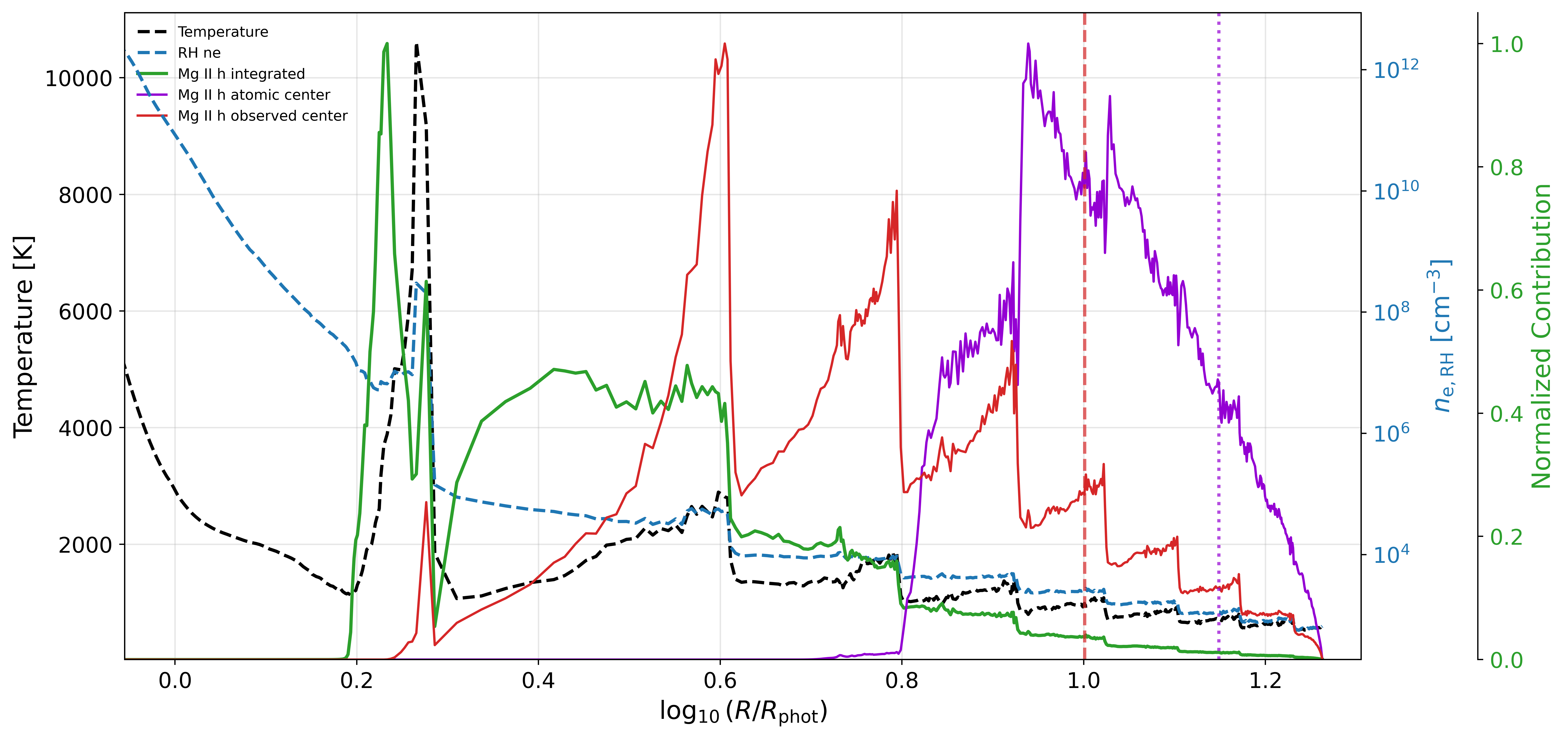}
  \caption{Contribution functions of the Mg\,\textsc{ii} k (top) and h (bottom) lines computed from the RH model atmosphere for P4. In each panel, the green curves show the wavelength-integrated contribution over the selected line window, the violet curve marks the contribution at the laboratory line center, and the red curve corresponds to the observed line-center wavelength. The black and blue dashed lines show the atmospheric temperature structure and the calculated electron density stratification as a function of radius. The vertical dotted and dash-dotted lines indicate the heights where the monochromatic optical depth reaches unity, $\tau_\nu=1$, for the laboratory and observed wavelengths, respectively; these heights provide an approximate reference for the main formation layers of the emergent radiation at those wavelengths. The comparison between the integrated and monochromatic contribution functions highlights the difference between the effective formation region of the full line profile and the more localized formation heights sampled by the line cores, illustrating the extended atmospheric layers probed by the Mg\,\textsc{ii} resonance lines.}
  \label{mgii-kh-contribution}
\end{figure*}

To investigate the atmospheric layers that contribute to the emergent Mg\,\textsc{ii} h \& k profiles and their relation to the temperature stratification, we calculated contribution functions from the RH output. Fig. \ref{mgii-kh-contribution} shows the normalized contribution as a function of atmospheric height for our main model using the P4 atmospheric structure.

To do this, we used the quantities given by RH at each wavelength and depth point: the source function $S_\nu$, the optical depth $\tau_\nu$, and the opacity $\chi_\nu$. From these, we calculated the contribution function as
\[
C_\nu(z) = S_\nu(z)\, e^{-\tau_\nu(z)}\, \chi_\nu(z),
\]
which describes the contribution of each atmospheric layer to the emergent intensity at a given wavelength. The factor $e^{-\tau_\nu}$ represents the attenuation by the material above, while $\chi_\nu$ weights the local emission by the opacity. For optically thick resonance lines such as Mg\,\textsc{ii} h \& k, the peak of the contribution function should therefore not be
interpreted simply as the location of maximum local emissivity, but rather as the region from which photons can most effectively escape after scattering.

For the Mg\,\textsc{ii} contribution-function analysis, we considered both the laboratory line centres, $\lambda_0$, and the observed peak wavelengths, $\lambda_{\mathrm{obs}}$, of the profiles as stated in the previous section.
To examine the combined contribution from the line core and wings, we integrated the contribution function over the range 2795--2797\,\AA\ for Mg\,\textsc{ii} and 2802--2804\,\AA\ for Mg\,\textsc{ii} h. Contributions from the line cores, as well as the integrated core and wing contributions for both Mg\,\textsc{ii} lines, are shown in Fig. \ref{mgii-kh-contribution}. The green, violet, and red curves represent the normalized contribution as a function of stellar radius for the wavelength-integrated range, at $\lambda_0$, and at $\lambda_{\rm obs}$, respectively. Fig. \ref{mgii-kh-contribution} also marks the heights where $\tau_\nu=1$ at $\lambda_0$ and $\lambda_{\rm obs}$, indicated by the vertical dotted violet and dashed red lines. These markers provide approximate escape heights for photons at those wavelengths, although the full line formation extends over a range of layers.

The contribution functions in Fig. \ref{mgii-kh-contribution} show that the Mg\,\textsc{ii} h \& k line-core photons escape from higher atmospheric layers than the wavelength-integrated core-plus-wing emission. This is expected for strong resonance lines with large line-centre opacity. When the contribution is integrated over a wider wavelength interval that includes both the core and the
wings, the effective contribution shifts toward deeper layers, because the line wings are less opaque and can sample denser material closer to the photosphere. Similar behaviour has been reported for the solar Mg\,\textsc{ii} h \& k lines \citep{Leenaarts2013,Magain1986}.

The Mg\,\textsc{ii} contribution functions also show that the emergent resonance-line radiation is produced and scattered over an extended region around and above the shock-heated layers. This is consistent with the interpretation of \citet{Wood2000,Wood2000ApJ}, in which Mg\,\textsc{ii}
photons are generated in the heated gas behind the outward-propagating shock, while the observed line profiles are subsequently modified by scattering and absorption in cooler material above the shock. Thus, the fact that the contribution-function peaks do not coincide exactly with the strongest temperature peak is not unexpected; for optically thick Mg\,\textsc{ii} resonance lines, the emergent profile is controlled not only by where photons are created, but also by where they last scatter and escape.

Moreover, looking at Fig. \ref{mgii-kh-contribution} suggest that the P4 atmospheric structure may not extend far enough to fully cover the outermost layers contributing to the Mg\,\textsc{ii} line cores, particularly for Mg\,\textsc{ii} k. To examine this, we extrapolated the atmospheric structure to $R/R_{\rm phot}=4$ and reran the model. In the extrapolated model, the
contributions of both Mg\,\textsc{ii} h and k decrease to zero by $R/R_{\rm phot}\simeq1.20$. Since the resulting contribution functions are almost identical to those shown in Fig. \ref{mgii-kh-contribution}, the extrapolated model is not shown.

\subsection{Comparison of Mg\,\textsc{ii} results with previous studies}\label{MgII-comparison}

The UV Mg\,\textsc{ii} h \& k resonance lines in Mira variables have been previously studied with \textit{IUE}. Previous studies showed that the Mg\,\textsc{ii} emission is strongly phase dependent: it typically appears after optical maximum, around $\phi\simeq0.1$, reaches maximum flux at
$\phi\simeq0.3$--$0.45$, and fades by $\phi\simeq0.7$ \citep{Brugel1986,Luttermoser1996,Wood2000}. The absolute Mg\,\textsc{ii} flux can also vary substantially from one pulsation cycle to another, in
some cases by 2--3 orders of magnitude \citep{Wood2000}. 
Additionally, Mg\,\textsc{ii} spectra in Mira variables show that the h \& k resonance lines have the largest blueshifts and broadest profiles among the UV emission lines \citep{Wood2000}. In their Mira sample, \citet{Wood2000} found that the Mg\,\textsc{ii} h line was consistently blueshifted, with
stellar-rest-frame velocities ranging from approximately $-70$ to $-40~{\rm km~s^{-1}}$ between $\phi\simeq0.2$ and $0.6$, while the corresponding FWHM values decreased from roughly $70$ to
$40~{\rm km~s^{-1}}$. They also showed that the magnitude of the blueshift and the line width both decrease with pulsation phase.

Our STIS observation of R~Leo was obtained at $\phi\simeq0.37$, close to the expected phase of maximum Mg\,\textsc{ii} emission, and the strong h \& k emission observed here is therefore consistent with previous phase-dependent studies.
The measured line parameters are listed in Table~\ref{MgII-table}. The Mg\,\textsc{ii} h line has a stellar-rest-frame velocity of $v_\star=-38.1~{\rm km~s^{-1}}$ and ${\rm FWHM}=44.5~{\rm km~s^{-1}}$, placing it near the lower end of the velocity and width ranges reported by \citet{Wood2000} for Mira variables at similar pulsation phases. The Mg\,\textsc{ii} k line has $v_\star=-28.7~{\rm km~s^{-1}}$, but its profile exhibits a double-peaked morphology, and its tabulated FWHM was measured from a Gaussian fit to the stronger component only rather than to the full profile. Combined with the likelihood of circumstellar absorption and self-reversal, this makes the
k-line centroid and width less reliable diagnostics than those of the h line.

The Mg\,\textsc{ii} h line is stronger than the k line in our R~Leo spectrum. Although the k line is usually stronger in optically thin conditions because of its larger oscillator strength, enhanced h/k ratios are common in Mira spectra \citep{Wood2000}. This behaviour is generally
attributed to stronger absorption and opacity effects in the k line, including overlying circumstellar Fe\,\textsc{i} and Mn\,\textsc{i} absorption near the Mg\,\textsc{ii} k wavelength.

The interpretation of the Mg\,\textsc{ii} blueshifts requires caution. As discussed by \citet{Wood2000ApJ}, the large velocity shifts observed in Mira variables do not necessarily trace the bulk motion of the shocked gas. Because Mg\,\textsc{ii} h \& k are optically thick resonance lines, their profiles are strongly influenced by high optical depth, resonance scattering, partial frequency redistribution, non-equilibrium ionization, self-absorption, velocity gradients, and absorption or scattering in material above the shock. The measured blueshifts and widths therefore reflect radiative-transfer
effects within a shock-structured atmosphere rather than the velocity of a single gas layer. The Mg\,\textsc{ii} h profile observed in R~Leo is consistent with this interpretation.

\subsection{NLTE calculations of C\,\textsc{ii}] multiplet}\label{CII-NLTE}

\begin{figure*}[]
 \centering
\includegraphics[width=0.85\textwidth]{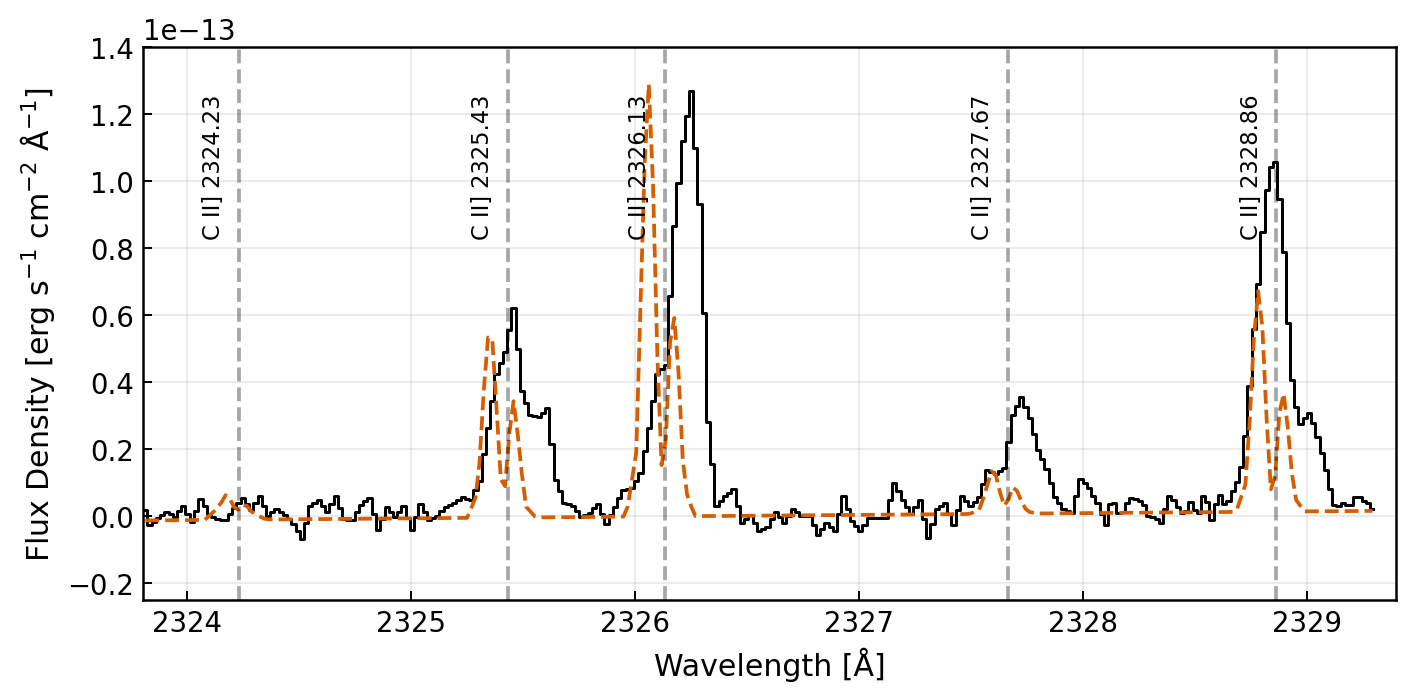}
 \caption[]{Observed C\,\textsc{ii}] multiplet lines toward R~Leo (black solid line), overlaid with the RH model at P4 (orange dashed lines).}
\label{CII-HST-RH}  
\end{figure*}

We performed RH non-LTE calculations for C, Mg, and H simultaneously using the P4 atmospheric structure. Fig.~\ref{CII-HST-RH} shows the RH model overlaid on the STIS spectrum of R~Leo in the region of the C\,\textsc{ii}] \(2325\,\AA\) multiplet. Overall, the model reproduces the observed C\,\textsc{ii}] multiplet reasonably well, including the relative strengths of the individual components.

Unlike Mg\,\textsc{ii}, the comparison between the RH calculations and the observed C\,\textsc{ii}] multiplet is more sensitive to the local continuum level. Although the absolute C\,\textsc{ii}] line fluxes can be approximately reproduced with an effective scaling radius of ($R_{\rm eff}=225,R_\odot$), this scaling places the RH continuum above the observed STIS continuum by ($5.2\times10^{-13},\mathrm{erg,s^{-1},cm^{-2},\AA^{-1}}$). We therefore subtracted this constant offset from the RH spectrum when producing Fig.~\ref{CII-HST-RH}, such that the model and observed continua coincide in the vicinity of the multiplet. Consequently, we regard ($R_{\rm eff}$) primarily as a normalization parameter used for comparing the model and observed fluxes rather than as a robust estimate of the projected emitting area of the C,\textsc{ii}] region.

The residual continuum mismatch is likely related to limitations of the adopted atmospheric model. In particular, uncertainties in the thermal and ionization structure of the layers where the C\,\textsc{ii}] multiplet forms may affect the predicted local UV continuum. In addition, the background opacity in this wavelength region may not be fully accounted for if relevant opacity sources are missing or imperfectly treated. These effects may influence the absolute continuum level without necessarily altering the overall quality of the fit to the emission lines. The contribution-function analysis provides a more reliable indication of the line-formation region and suggests that the C\,\textsc{ii}] emission originates in a relatively narrow shell-like or localized region of the atmosphere.

\begin{figure*}[]
 \centering
\includegraphics[width=0.85\textwidth]{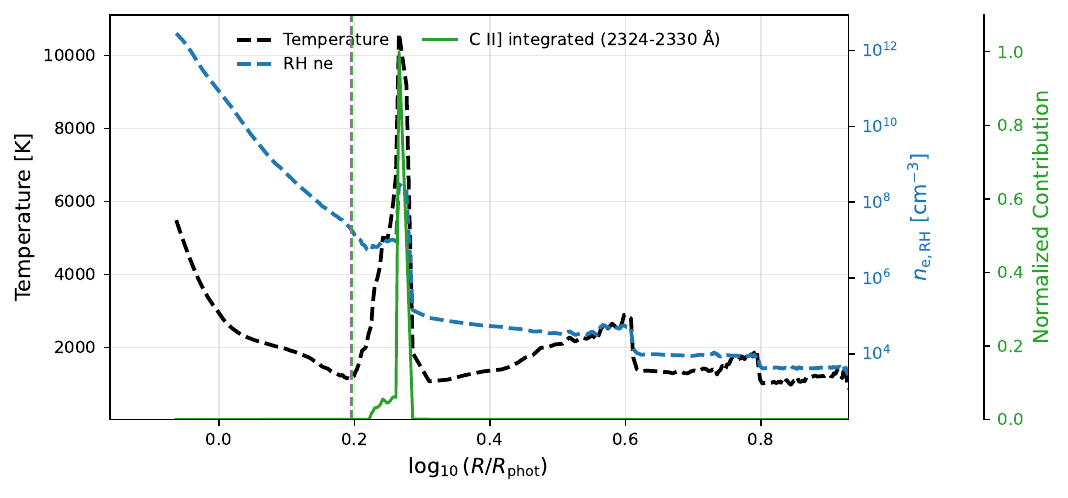}
\caption[]{Radial formation diagnostics for the C\,\textsc{ii}] multiplet in the RH model atmosphere. The black dashed curve shows the temperature structure, the blue dashed curve the RH-computed electron density, and the green curve the band-integrated C\,\textsc{ii}] contribution function, normalized to its own peak. The C\,\textsc{ii}] emission is strongly localized in a narrow shell at \(\log_{10}(R/R_{\rm phot}) \approx 0.26\), coincident with the sharp temperature rise to \(10610\,\mathrm{K}\) and elevated electron density of order \(3\times10^{8}\,\mathrm{cm^{-3}}\).}
\label{CII-contribution}  
\end{figure*}

Figure~\ref{CII-contribution} presents the normalized contribution of the integrated C\,\textsc{ii}] multiplet over the wavelength range \(2324\)--\(2330\,\AA\), including all five components, as a function of atmospheric height. The figure also shows the temperature profile and the RH-computed electron density, illustrating the relation between the atmospheric structure and the formation of the C\,\textsc{ii}] emission.
As shown in Fig.~\ref{CII-contribution}, the C\,\textsc{ii}] \(\lambda 2325\) multiplet forms predominantly in a compact hot outer layer rather than throughout the extended atmosphere. The normalized contribution profile peaks at \(R/R_{\rm phot}\approx 1.84\), where the atmosphere undergoes a sharp temperature rise and the RH-computed electron density remains relatively high. At this radius, the local conditions are \(T\approx 1.06\times10^{4}\,\mathrm{K}\) and \(n_{\mathrm e}\approx 3.0\times10^{8}\,\mathrm{cm^{-3}}\). In contrast, the line-center \(\tau_\nu=1\) surfaces for the five components lie deeper in the atmosphere, at \(R/R_{\rm phot}\approx 1.57\). The narrowness of the contribution profile shows that the emission arises in a geometrically localized shell rather than throughout the extended atmosphere as a whole. Moreover, all five C\,\textsc{ii}] components remain optically thin at the radius where their contribution functions peak, with \(\tau_\nu\approx 0.08\)--0.13. This demonstrates that the C\,\textsc{ii}] emission does not form primarily at the \(\tau_\nu=1\) depth, but instead in a hotter, optically thin outer layer.

As discussed in Sect. \ref{CII-Ne-results}, an independent estimate of the electron density was obtained from the observed C\,\textsc{ii}] line ratio R1, R2, and R3. Using CHIANTI, this ratio yields \(n_{\mathrm e}\approx 8.9\times10^{8}\,\mathrm{cm^{-3}}\). This value is somewhat higher than the RH-computed local density at the peak of the C\,\textsc{ii}] contribution function, \(n_{\mathrm e}\approx 3.0\times10^{8}\,\mathrm{cm^{-3}}\). However, a more rigorous comparison with the RH model can be made by computing a contribution-weighted electron density over the specific line-forming region of the two \(R_3\) diagnostic lines,
\[
\langle n_{\mathrm e}\rangle_{R_3}
=
\frac{\int n_{\mathrm e}(r)\,[C_{2325.430}(r)+C_{2327.665}(r)]\,dr}
{\int [C_{2325.430}(r)+C_{2327.665}(r)]\,dr},
\]
which gives \(\langle n_{\mathrm e}\rangle_{R_3}\approx 1.9\times10^{8}\,\mathrm{cm^{-3}}\). The CHIANTI-based density is therefore higher than the RH peak density by a factor of \(\sim 2.9\), and higher than the contribution-weighted RH density by a factor of \(\sim 4.7\).

This difference is not necessarily unexpected. The CHIANTI analysis assumes a single optically thin plasma characterized by one representative temperature and density, whereas the RH calculation predicts line formation in a stratified atmosphere with substantial gradients in temperature and electron density. The line-ratio estimate should therefore be interpreted as an effective single-zone density, rather than as a precise measurement of the local electron density at the depth of maximum line formation. In this sense, the RH model provides a more physically detailed description of the C\,\textsc{ii}] emitting region, while the CHIANTI result remains a useful first-order diagnostic.

The contribution-function analysis therefore supports a picture in which the C\,\textsc{ii}] multiplet originates in a narrow shell in the outer atmosphere, where the temperature rises steeply and the electron density remains sufficiently high to produce the observed semi-forbidden emission.

\subsection{Kinematics of the shocked material}\label{Outflow}

The Al\,\textsc{ii}] $\lambda2669$ (UV1) line provides an additional kinematic constraint on the shocked material because, as a semi-forbidden transition, it is expected to be much less affected by opacity and resonant scattering than the Mg\,\textsc{ii} h \& k lines \citep[e.g.][]{Wood2000}. The fitted centroid, $\lambda_{\rm c}=2669.9612$~\AA, corresponds to a heliocentric velocity of $v_{\rm obs}=1.14\pm0.12~{\rm km~s^{-1}}$. After correcting for the adopted center-of-mass velocity of R~Leo, $V_{\rm CoM}=7.2~{\rm km~s^{-1}}$, this gives a stellar-rest-frame velocity of $v_\star \simeq -6.1~{\rm km~s^{-1}}$.

This projected velocity is smaller than, but comparable in magnitude to, the $10.6\pm1.4~{\rm km~s^{-1}}$ outward expansion of the millimetre-wavelength surface of R~Leo measured with ALMA by \citet{Vlemmings2019}. The comparison should be interpreted cautiously, since the Al\,\textsc{ii}] centroid measures a line-of-sight Doppler velocity of the UV-emitting gas, whereas the ALMA value traces the apparent radial expansion of an optically thick continuum surface. Thus, an outward-moving layer on the near side of the star would naturally appear blueshifted in the UV line.

We note that the inferred Al\,\textsc{ii}] stellar-rest-frame velocity depends directly on the adopted systemic velocity of R~Leo, which is uncertain for pulsating Mira variables. Nevertheless, our derived Al\,\textsc{ii}] velocity is consistent with the weighted-average Al\,\textsc{ii}] $\lambda2669$ velocity reported for R~Leo by \citet{Wood2000}. It is also consistent with expected shock speeds in Mira atmospheres, which are typically of order $10$--$20~{\rm km~s^{-1}}$ \citep{Bowen1988,Gillet1988}.

The Al\,\textsc{ii}] velocity is also comparable to the modest velocities measured for the Mg\,\textsc{ii} UV2 and UV3 lines and the cleaner C\,\textsc{ii}] multiplet components, in contrast to the much larger blueshifts of the optically thick Mg\,\textsc{ii} h \& k resonance lines.

The measured Al\,\textsc{ii}] FWHM is $0.2119$~\AA, corresponding to an observed line width of $23.80\pm0.33~{\rm km~s^{-1}}$. This width represents the line broadening within the emitting region and may include turbulent broadening, thermal broadening, and unresolved velocity structure along the line of sight.

\begin{figure}[]
 \centering
\includegraphics[width=0.48\textwidth]{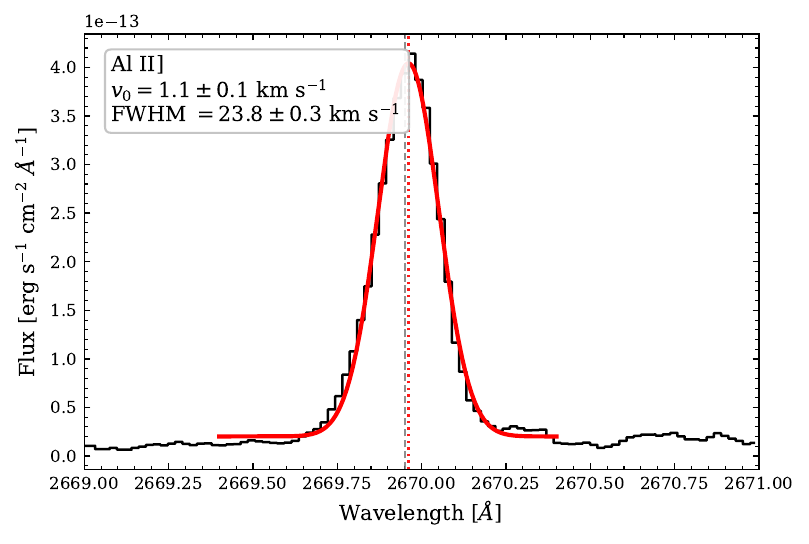}
 \caption[]{Observed Al\,\textsc{ii}] $\lambda2669$ line profile toward R~Leo. 
    The red curve shows a Gaussian fit to the line profile, and the vertical dashed line marks the laboratory rest wavelength. The fitted centroid corresponds to a heliocentric velocity of $v_{\rm obs}=1.14~{\rm km~s^{-1}}$. After subtracting the  $v_{\rm CoM}=7.2~{\rm km~s^{-1}}$, the Al\,\textsc{ii}]-emitting gas is blueshifted by $-6~{\rm km~s^{-1}}$.}
\label{Al}  
\end{figure}

\section{Summary}\label{Summary}

An important open question in stellar astrophysics is which physical processes heat stellar chromospheres during different phases of stellar evolution. In evolved stars this problem is particularly challenging because pulsation, shocks, mass loss, circumstellar material, and possible magnetic activity can affect many of the same observables. In AGB stars, recent high-resolution mm/sub-mm observations have revealed asymmetric extended atmospheres and hot
compact structures, but the connection between these features and the ultraviolet chromospheric diagnostics remains poorly understood.

In this work, we analysed high-resolution STIS/\textit{HST} ultraviolet spectra of the Mira variable R~Leo. The spectrum contains emission from thirteen atomic species, including Mg\,\textsc{ii}, C\,\textsc{ii}], Fe\,\textsc{ii}, and Al\,\textsc{ii}]. We focused on the Mg\,\textsc{ii} h \& k resonance lines and the C\,\textsc{ii}] 2325\,\AA\ multiplet, and modelled their formation with the NLTE radiative-transfer code RH using phase-dependent hydrodynamic atmospheric structures that include pulsation-driven shocks. A model atmosphere close to pulsation phase $\phi\simeq0.4$ provides a reasonable match to the spectrum
observed at $\phi\simeq0.37$, indicating that pulsation-driven atmospheric structure plays a major role in shaping these UV diagnostics.

The C\,\textsc{ii}] multiplet provides an electron-density diagnostic for the shock-heated gas. From the observed multiplet ratios, we estimate an electron density of order of $n_e\simeq10^9\,\mathrm{cm^{-3}}$. In this analysis, the C\,\textsc{ii}] 2326.133\,\AA\ component was corrected for flux lost through Fe\,\textsc{i} pumping/fluorescence. The RH contribution functions show that the C\,\textsc{ii}] emission arises mainly from a compact shock-heated shell at $R/R_{\rm phot}\approx1.84$, where the local temperature is about $10^4$\,K and the RH-computed electron density is of order $3\times10^8\,\mathrm{cm^{-3}}$.

The Mg\,\textsc{ii} h \& k lines probe a more extended atmospheric region than the C\,\textsc{ii}] multiplet. The contribution functions indicate that the Mg\,\textsc{ii} cores form at larger radii than the wings, broadly analogous to the height-dependent formation of strong chromospheric lines in the Sun. The Mg\,\textsc{ii} h line is reproduced more reliably than the k line, which is more strongly affected by circumstellar absorption and opacity effects. The strong phase dependence of the synthetic Mg\,\textsc{ii} profiles further
supports the interpretation that pulsation-driven shocks strongly influence the
UV chromospheric emission.

The semi-forbidden Al\,\textsc{ii}] $\lambda2669$ line provides an additional low-opacity kinematic diagnostic. Its centroid indicates a projected stellar-rest-frame blueshift of about $6~{\rm km~s^{-1}}$, smaller than but comparable in magnitude to the $10.6\pm1.4~{\rm km~s^{-1}}$ outward expansion of the millimetre-wavelength surface of R~Leo measured with ALMA. This supports a consistent picture in which UV semi-forbidden emission and mm/sub-mm
continuum variability both trace shock-structured material in the extended atmosphere.

Future progress will require time-monitoring observations that combine UV, optical, and mm/sub-mm diagnostics over the pulsation cycle, together with detailed NLTE modelling for a larger sample of AGB stars. Such observations will be essential for connecting chromospheric UV emission, shock propagation, atmospheric asymmetries, and the onset of mass loss in cool evolved stars.

\begin{acknowledgements}
This research is based on observations made with the NASA/ESA Hubble Space Telescope obtained from the Space Telescope Science Institute, which is operated by the Association of Universities for Research in Astronomy, Inc., under NASA contract NAS 5--26555. These observations are associated with program 7756.
This work is supported by the ESGC project (project No. 335497) funded by the Norwegian Research Council.

\end{acknowledgements}




\bibliographystyle{aa} 
\bibliography{refrences} 


\begin{appendix} 

\section{Bowen atmospheric structure}

\begin{figure*}
  \centering
  \includegraphics[width=0.89\textwidth]{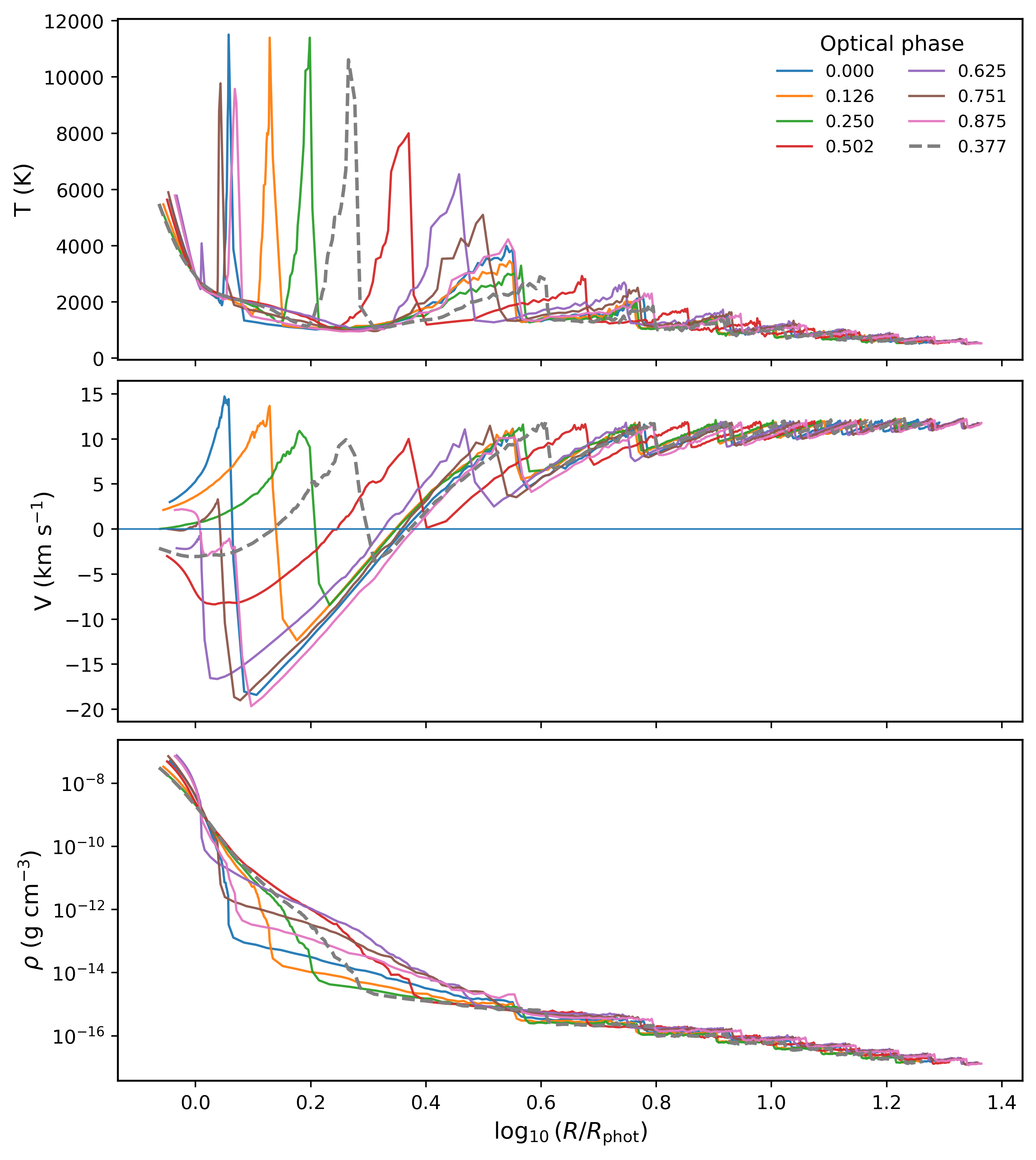}
  \caption{Gas temperature, velocity, and density stratifications for all phases of a Mira pulsation cycle from the Bowen (1988) dynamical model as a function of radius. The phase 0.377, corresponding to the observational epoch, is indicated by dashed lines.}
  \label{Fig-Bowen-8phase}
\end{figure*}

Figure~\ref{Fig-Bowen-8phase} shows the atmospheric stratification for all eight phases of the hydrodynamic models presented by \cite{Bowen1988}, illustrating how the atmospheric structure evolves as the pulsation-driven shock propagates outward through the stellar atmosphere. The size of photosphere in all phases are also listed in Table \ref{Bowen-photosphere}.

\begin{figure*}
  \centering
  \includegraphics[width=\textwidth]{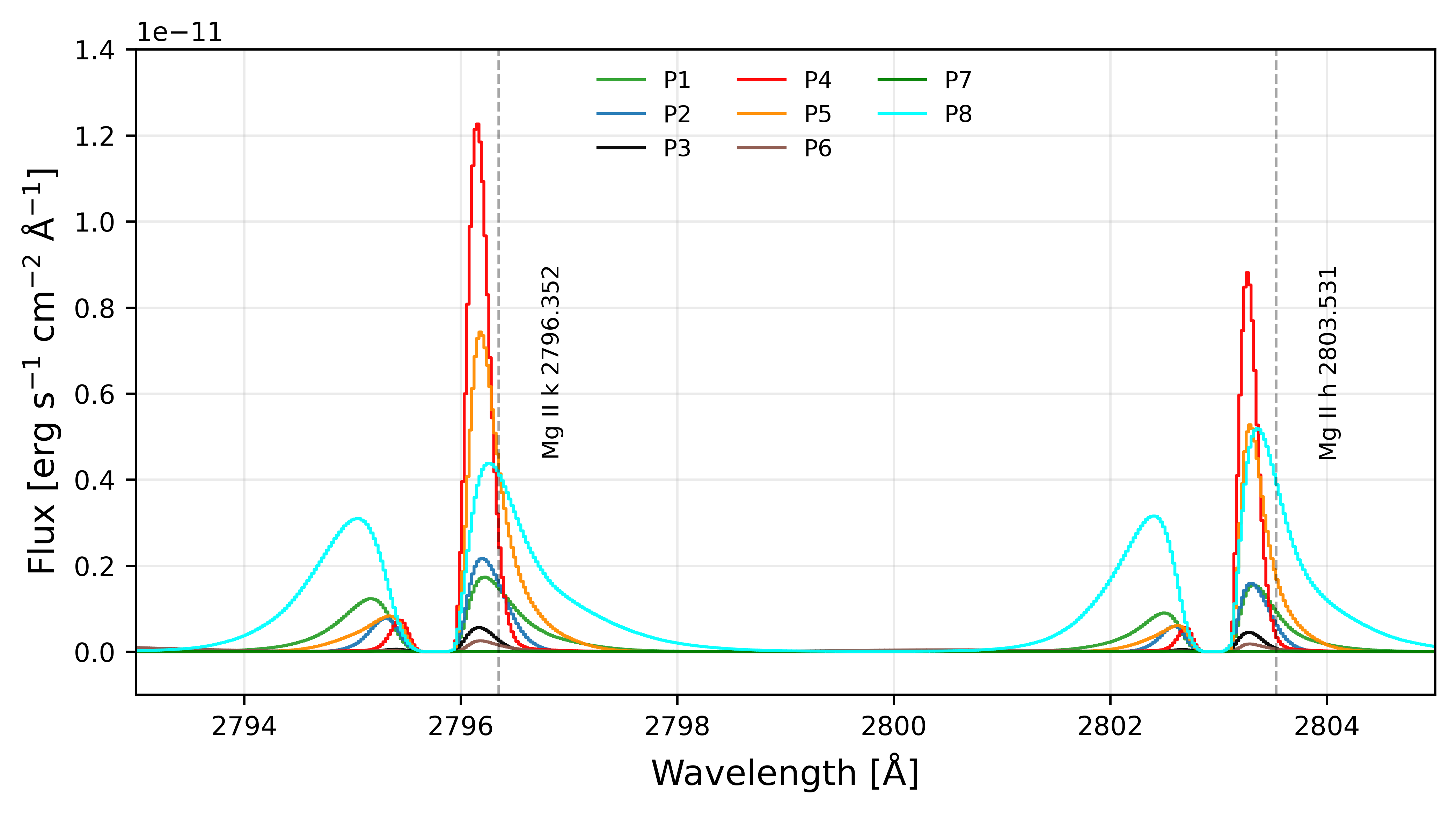}
  \caption{RH model predictions of Mg\,\textsc{ii} $h$ \& $k$ for Bowen phases P1--P8. The synthetic spectra are scaled to an angular diameter corresponding to 110 $R_{\odot}$ for all phases; however, their effective sizes may differ as the photospheric radius varies, as listed in Table~\ref{Bowen-photosphere}. The figure is intended to illustrate the dependence of the Mg\,\textsc{ii} $h$ \& $k$ line-profile variations on pulsation phase, rather than to compare absolute fluxes.}
  \label{Fig-MgII-8phase}
\end{figure*}

Figure~\ref{Fig-MgII-8phase} shows the RH model results for the Mg\,\textsc{ii} $h$ \& $k$ lines using Bowen atmospheric stratification at eight Bowen pulsation phases shown in Fig.~\ref{Fig-Bowen-8phase} as input. We note that all spectra scaled to the same angular diameter corresponding to 110$R_{\odot}$ although the photospheric radius varies over the pulsation and the effective emitting region may also vary. Therefore, the figure should be interpreted mainly in terms of line-profile morphology and phase-dependent variability, rather than absolute flux differences. The synthetic profiles show a strong dependence on pulsation phase. Some phases produce very weak or nearly absent Mg\,\textsc{ii} emission, while others show prominent emission with different widths, peak positions, and wing extensions. In particular, phases around P4--P5 produce the strongest narrow core emission in this scaling, whereas P8 shows broader profiles with more extended wings. These changes reflect the evolution of the temperature, density, and velocity structure as the pulsation-driven shock propagates through the atmosphere. The comparison demonstrates that the Mg\,\textsc{ii} $h$ and $k$ line shapes are highly sensitive to the pulsation phase and can therefore be used as diagnostics of the time-dependent atmospheric structure.

\begin{table}
\centering
\caption{Optical phases selected from the Bowen model and the corresponding photospheric radii.}
\label{Bowen-photosphere}
\setlength{\tabcolsep}{2.4pt}
\begin{tabular}{l|cccccccc}
\hline
ID & P1 & P2 & P3 & P4 & P5 & P6 & P7 & P8 \\
\hline
\tiny Optical phase & 0.0 & 0.126 & 0.250 & 0.377 & 0.502 & 0.625 & 0.751 & 0.875 \\
$R_{\rm phot}$ (\tiny $10^{13}$\,cm) & 1.663 & 1.813 & 1.880 & 1.843 & 1.679 & 1.517 & 1.526 & 1.528 \\
\hline
\end{tabular}
\end{table}

\end{appendix}
\end{document}